\documentclass[%
reprint,
superscriptaddress,
frontmatterverbose, 
nofootinbib,
amsmath,amssymb,
aps,
prstab,
prstper,
floatfix,
]{revtex4-2}

\usepackage{longtable}
\usepackage{graphicx}
\usepackage{dcolumn}
\usepackage{bm}
\usepackage{hyperref}
\usepackage[mathlines]{lineno}
\usepackage[caption=false]{subfig}
\usepackage{ulem} 
\usepackage{comment}
\usepackage{orcidlink}

\usepackage{array} 

\usepackage[
  scale=0.7,
  marginratio={1:1, 2:3},
  ignoreall,
  text={7in,10in},
  centering,
  margin=0.24in,
  total={6.5in,8.75in},
  top=1.2in,
  left=0.6in,
  right=0.6in,
  includefoot
]{geometry}

\usepackage{threeparttable}
\usepackage{setspace}
\hypersetup{colorlinks=true, linkcolor=blue,      
    urlcolor=blue, 
	citecolor=green,  
    linktocpage =true}
\usepackage{url}
\usepackage{multirow}
\usepackage{standalone}
\usepackage{cancel}
\usepackage{ulem}
\usepackage{colortbl}
\usepackage{enumitem}
\usepackage{tablefootnote}

\usepackage{color}
\usepackage{natbib}
\usepackage{epsfig}
\usepackage{float}
\usepackage{xcolor}
\usepackage{booktabs}
\usepackage{bibunits}
\usepackage{amsmath,amssymb,amsfonts}

\begin{document}

\preprint{APS/123-QED}

\title{Robustness of Sensitivity Evaluations for Gravitational Wave Detection Algorithms}

\author{Alexandra E. Koloniari}
\affiliation{Department of Physics, Aristotle University of Thessaloniki, 54124 Thessaloniki, Greece}

\author{Lazaros Lazaridis}
\affiliation{Department of Physics, Aristotle University of Thessaloniki, 54124 Thessaloniki, Greece}

\author{Christos Paschalidis}
\affiliation{Department of Physics, Aristotle University of Thessaloniki, 54124 Thessaloniki, Greece}

\author{Nikolaos Stergioulas}
\affiliation{Department of Physics, Aristotle University of Thessaloniki, 54124 Thessaloniki, Greece}

\date{\today}

\begin{abstract}
The discovery of gravitational waves (GWs) from merging compact binaries has transformed modern astrophysics, driving innovation in detection methodologies. Whereas matched-filtering techniques have long been the standard, the growing volume of data from advanced observatories like LIGO, Virgo, and KAGRA has spurred interest in machine learning (ML) solutions for their scalability and computational efficiency. As next-generation detectors approach reality, the development of reliable and adaptable search algorithms becomes increasingly urgent. This work examines the consistency of detection sensitivity in AresGW model 1, an ML-based pipeline, when applied to multiple month-long datasets consisting of real detector noise. By analyzing the number of waveform injections detected and the sensitive distance at different FAR thresholds, we evaluate the model's performance under low false alarm rates and investigate how sensitivity metrics fluctuate due to dataset variability. In addition, we evaluate the performance of our algorithm on data both with and without contamination of genuine GW signals. Our findings reveal notable performance variations, highlighting the challenges introduced by finite-duration datasets and emphasizing the need for more rigorous statistical validation. By identifying these challenges, we aim to clarify the practical limitations of both ML-based and traditional detection systems and inform future benchmarking standards for GW searches.
\end{abstract}

\maketitle

\section{\label{sec:intro} Introduction}

The direct detection of gravitational waves (GWs) from compact binary coalescences (CBCs) has become a cornerstone of modern astrophysics. To enable such detections, a diverse suite of search algorithms has been developed. These range from traditional matched-filtering pipelines — such as PyCBC \cite{usman2016pycbc, alex_nitz_2022_6912865, 2024arXiv240310439K}, MBTA \cite{2021CQGra..38i5004A} and GstLAL\cite{gstlal, gstlal2} — to wavelet-based searches like cWB \cite{PhysRevD.93.042004,cwb-softwareX, klimenko_sergey_2021_5798976}, and more recently to data-driven approaches leveraging deep learning techniques \cite{cuoco2020review,app13179886,2023arXiv231115585Z}.

In particular, the success of the ongoing fourth observing run (O4) of the LIGO-Virgo-KAGRA collaboration \cite{LIGO_ref, Virgo_ref, KAGRA_ref} and the increased computational requirements of future runs have accelerated the adoption of machine learning (ML) in the analysis of GW data. This shift is largely motivated by the computational challenges inherent in conventional matched-filtering approaches. Thus, a variety of ML methods — including convolutional neural networks (CNNs), autoencoders, and other deep learning architectures — have been explored and evaluated in the last few years \cite{PhysRevLett.120.141103, PhysRevD.97.044039, 2019PhRvD.100f3015G, CORIZZO2020113378, PhysRevD.102.063015, 2020PhRvD.101j4003W, 2020PhLB..80335330K, 2020arXiv200914611S, 2021PhRvD.103f3034L, DODIA, 2021NatAs...5.1062H, 2021MNRAS.500.5408M, 2021PhLB..81236029W, 2021CQGra..38o5010A, 2021PhRvD.104f4051J, 10.3389/frai.2022.828672, 2022arXiv220208671C, PhysRevD.105.043003, 2022arXiv220606004B, schafer2022training, PhysRevD.106.042002, 2022arXiv220704749A, 2022arXiv220612673V, 2022PhRvD.106h4059A, 2022MNRAS.516.3847G, 2023PhRvD.107h2003A, 2023MNRAS.519.3843A, 2023PhRvL.130q1402L, 2023PhRvL.130q1403D, 2023CQGra..40m5008B, 2023arXiv230615728T, 2023PhRvD.108d3024M, 2023MLS&T...4c5024B, 2022arXiv220111126M, 2023PhLB..84037850Q, 2023arXiv230200295L, 2023arXiv230519003J, 2023CQGra..40s5018F, 2023arXiv230716668G, 2023PhRvD.107d4032Y, 2023arXiv230808429B, 2023MLS&T...4d5028J, 2024ChJPh..88..301T, 2024arXiv240318661M, 2024PhRvD.109d3011S, 2024arXiv240207492Z}.  Notably, recent ML-driven breakthroughs include the detection of a new candidate event in a single-detector search of O3 data \cite{2024CQGra..41l5003T}, and the identification of eight additional candidates using the ResNet-based pipeline AresGW model 2 for a network of detectors using O3 data \cite{Koloniari2025}.

Looking ahead, the need for efficient and scalable detection algorithms will only grow with the advent of next-generation GW detectors - such as LIGO-India \cite{LIGO_India_2022}, Voyager \cite{LIGO:2020xsf}, Virgo nEXT, Cosmic Explorer \cite{2019BAAS...51g..35R}, Einstein Telescope \cite{2020JCAP...03..050M} and NEMO \cite{Ackley:2020atn}. These facilities will provide unprecedented sensitivity, amplifying both signal volume and noise complexity, thus reinforcing the relevance of ML-based approaches.

Thus, to understand how we arrived at this point and also how to navigate the path ahead, we will first examine the origins of the field, since as mentioned above, machine learning in GW detection has progressed significantly over recent years. One of the major milestones in assessing the readiness of ML approaches for real-world application was the first Machine Learning Gravitational-Wave Mock Data Challenge (MLGWSC-1) \cite{challenge1}. The challenge provided a common benchmark for evaluating the sensitivity and efficiency of ML models relative to traditional algorithms, using simulated injections into both Gaussian and real O3a detector noise. The real noise used was cleaned of events from GWTC-2 \cite{GWTC2}; however, events from the AresGW \cite{Koloniari2025}, GWTC-2.1 \cite{LIGOScientific:2021usb}, IAS \cite{IAS_O3a, IAS_higher_harmonics}, and OGC \cite{OGC-3, OGC-4} catalogs remained present in the noise data.

Participants were evaluated on 1-month datasets, with the sensitive distance at specific false alarm rate (FAR) thresholds and the runtime as  criteria, allowing for controlled yet realistic comparisons. The results of the challenge revealed a wide spread in algorithm performance -- particularly in the dataset with real noise -- suggesting that conclusions about model efficacy were conditionally valid. In the most demanding dataset with real O3 noise, the leading algorithm was Virgo-AUTH, which was the predecessor of AresGW model 1 \cite{AresGW_model,AResWG}, the model we use in the present analysis.

Using the sensitive distance as a metric, AresGW model 1 was the first machine learning model to achieve a higher sensitivity than a specific matched-filtering configuration\footnote{PyCBC participated using a standard configuration, not particularly optimized for the mass range of the test data in MLGWSC-1.}, marking a significant milestone. However, due to the limited duration of the dataset and the closeness of performance between AresGW model 1 and PyCBC, no percentage-based performance gap was reported to avoid overstating significance. Instead, the key takeaway was that AresGW model 1 can exceed traditional methods under the specific conditions of the challenge. Furthermore, in the same work, to account for statistical variability in performance, the variance of the AresGW model across different one-month test datasets was also reported.  This variance was at the level of 3\% at FAR of 1/month (and smaller at higher FAR).

A recent study \cite{nagarajan2025identifyingmitigatingmachinelearning} reanalyzed the MLGWSC-1 dataset using real noise, comparing a new machine learning-based detection pipeline with several existing GW search methods, including AresGW model 1, at various FARs. The work also examined common sources of bias in ML and adapted these insights to the GW detection domain, proposing mitigation strategies. Notably, it presented the number of detections at fixed FAR as an additional metric alongside the more commonly used sensitive distance, noting that the former can hide biases while the latter can be skewed by its dependence on chirp mass. As we will demonstrate, the number of detections at fixed FAR can indeed exhibit systematic bias, whereas sensitive distance proves to be more robust — particularly in terms of variability across datasets.

In addition, while the study presented in \cite{nagarajan2025identifyingmitigatingmachinelearning} offers performance comparisons, its reliance on the single month of the MLGWSC-1 test data raises concerns about statistical uncertainty, when comparisons are presented in terms of the number of injections at FAR of 1/month.
The noise background in GW detectors is known to vary significantly over time due to environmental and instrumental factors. Consequently, sensitivity metrics — such as the number of injections detected at a fixed FAR — can fluctuate across different datasets, even when the injection set remains constant. In this work, we will explore the variability of this metric, as well as the variability of the sensitive distance across different datasets.

Moreover, including real GW signals in the evaluation dataset (as was e.g. the case in the MLGWSC-1 dataset \cite{challenge1}, where only the GWTC-2 \cite{GWTC2} events were excluded from the test set), can affect the accuracy of background estimation, potentially leading to artificially reduced performance metrics for pipelines optimized to identify sub-threshold events. Some of the real GW signals included in that evaluation set had originally been identified by the improved version of AresGW model 1, AresGW model 2 \cite{Koloniari2025}.  Careful curation of evaluation data is, therefore, critical for obtaining reliable performance assessments, as will be demonstrated here.

Thus, in this work, we present a systematic study of detection sensitivity variance for the AresGW model 1 across multiple independent one-month datasets. Our goal is to assess the stability of two performance metrics, the number of detected injections and the sensitive distance, under conditions of realistic detector noise, and to quantify the effect of dataset choice — especially at low FAR thresholds. In addition, we address the issue of data contamination by real signals and its immediate effect on detection performance. Hence, we analyze injection recovery as a function of chirp mass ($\cal M$) and FAR, both with and without contamination by real events, by computing the confidence intervals for the mean, the standard deviation and the coefficient of variation in three different dataset categories: Datasets with identical noise and varying injections, Datasets with varying noise and identical injections, and Datasets with both varying noise and injections. Furthermore, we also conduct the same analyses using the sensitive distance at specific FAR thresholds as a metric. In doing so, our objective is to assess the robustness of both metrics and to provide a statistically grounded understanding of the capabilities and limitations of the model, while offering guidance for robust benchmarking practices in future traditional and machine learning-driven GW searches.

For clarification, we note that we chose to use AresGW model 1 for this analysis, despite the existence of the newer AresGW model 2 \cite{Koloniari2025}, for several reasons. First, employing model 1 allows for a direct comparison with the analysis presented in \cite{nagarajan2025identifyingmitigatingmachinelearning}. Second, AresGW model 1 is publicly available in \cite{AResWG}, facilitating the reproducibility of our results. Finally, AresGW model 2 includes three distinct output classes, which would require separate analyses for each class — introducing additional complexity beyond the intended scope of this study.

The paper is organized as follows. In Sec. \ref{sec:AresGW_model1}, we describe the architecture and training methodology of AresGW model 1. Sec. \ref{sec:test_datasets} outlines the construction and categorization of the test datasets used in our analysis. In Sec. \ref{sec:performance}, we evaluate the model’s performance on datasets containing known gravitational-wave events. Sec. \ref{sec:performance_clean} extends this analysis to datasets that have been cleaned of all known GW signals. In Sec. \ref{sec:fair_comparisons}, we discuss best practices and methodological recommendations for the robust evaluation of GW detection algorithms. Finally, we summarize our findings and implications in Sec. \ref{sec:results}.

\section{\label{sec:AresGW_model1} 
AresGW Model 1 Architecture and Training} 

The AresGW model 1 algorithm builds on the foundation of deep residual networks (ResNets \cite{wightman2021resnet}), using a 54-layer architecture enhanced with deep adaptive input normalization (DAIN \cite{dain}) to handle the complexities of real, non-stationary noise from LIGO’s O3 data. Although it maintains the same depth, 54 layers, as the previous model (Virgo-Auth model \cite{challenge1}), it doubles the number of filters, improving its feature extraction capabilities, just like the later version of the code, AresGW model 2 \cite{Koloniari2025}. 

The training process was completed in 31 hours over 14 epochs, while the evaluation on a month-long test dataset took approximately 2 hours — 1 hour for foreground and 1 hour for background evaluation. Training was performed on an A6000 GPU using a 12-day dataset of simulated BBH waveforms, generated with the IMRPhenomXPHM waveform model~\cite{IMRPhenomXPHM_ref}, with component masses between $7\,M_\odot$ and $50\,M_\odot$. The spins follow an isotropic distribution with magnitudes between 0 and 0.99, and the orientations are chosen randomly. Furthermore, the waveforms are uniformly distributed across various parameters, including coalescence phase, polarization, inclination, declination, and right ascension. These waveforms are embedded in real LIGO O3a noise, selected from data-quality segments where both LIGO detectors were operational and each segment was at least two hours in duration. Segments from the Livingston detector were previously randomly time-shifted and shuffled to ensure they could be distinguished from real coincident data. For additional details, see \cite{challenge1, AresGW_model}.

\begin{figure}[t]
  \centering
  \includegraphics[width=0.99\linewidth]{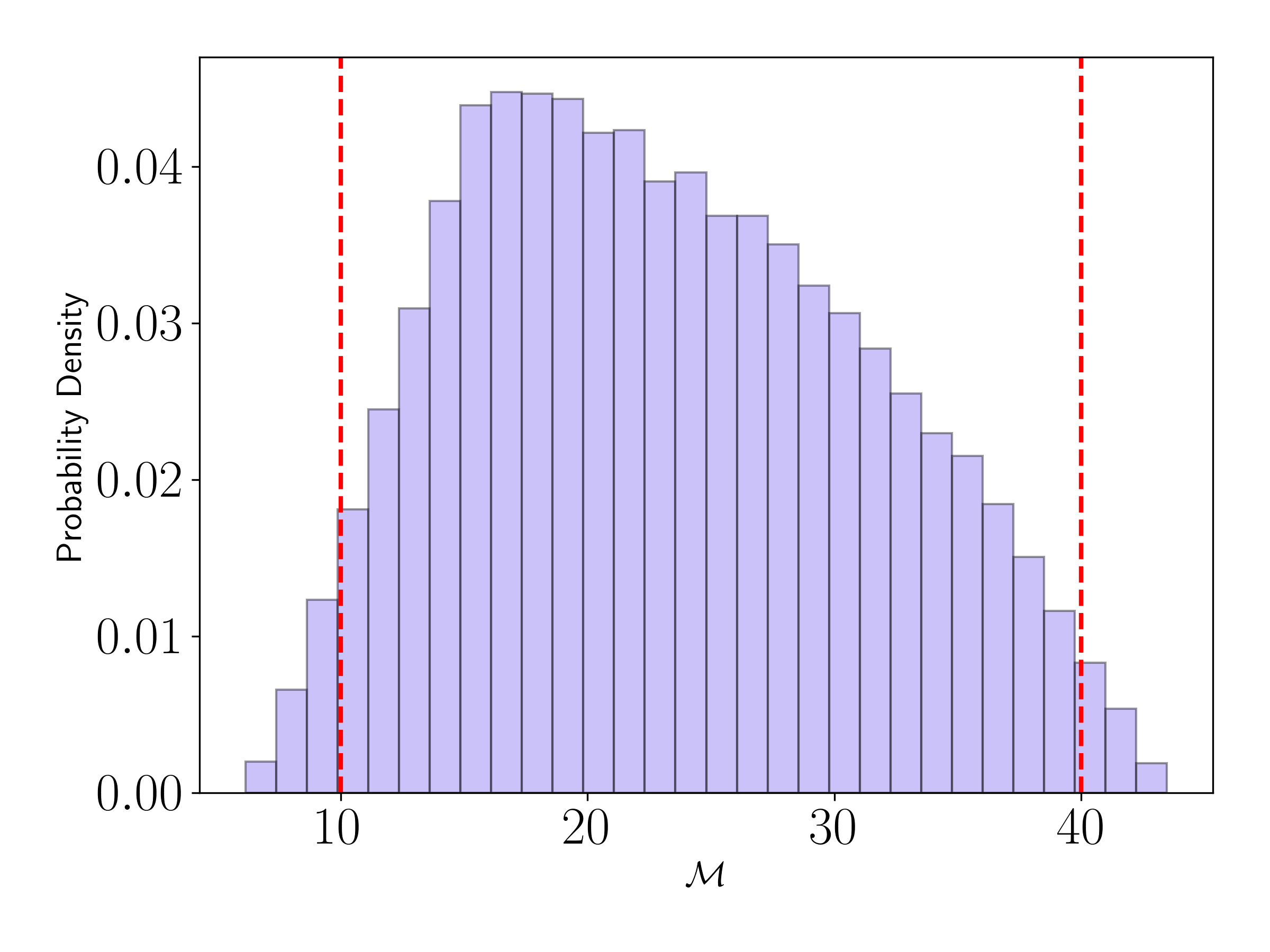}
    \caption{Chirp mass (${\cal M}$) distribution of the training and test sets. Although the individual masses $m_1$ and $m_2$ are uniformly distributed, they enter nonlinearly in  ${\cal M}$, resulting in a skewed ${\cal M}$ distribution. Only ~2.7\% of training waveforms have ${\cal M} < 10$ and ~1.7\% have ${\cal M} > 40$. The dashed red lines at ${\cal M} = 10$ and ${\cal M} = 40$ mark the effective training range, beyond which the network lacks sufficient exposure.}
    \label{fig:Mchirp_hist}
\end{figure} 

Note that the selection of individual source masses, as just outlined, leads to the $\mathcal{M}$ distribution illustrated in Fig.\ref{fig:Mchirp_hist}. This distribution has a mean of 23.2$M_\odot$ and a variance of 8.08~$M_\odot$. Apparently, although the component masses were drawn uniformly between 7 and 50~$M_\odot$, the resulting $\mathcal{M}$ values, are not evenly distributed due to their nonlinear relationship with the component masses. Instead, $\mathcal{M}$ exhibits a beta-like, skewed distribution, with approximately 95\% of values concentrated between 10 and 40~$M_\odot$. Values of $\mathcal{M}$ outside this range lie beyond the regime in which the model has been effectively trained, which could lead to reduced sensitivity to such signals. For more information on this topic, see \cite{Koloniari2025}.

Furthermore, unlike the Virgo-AUTH model, AresGW model 1 incorporates dynamic dataset augmentation (random noise segments are dynamically swapped with different detector data during training to enhance generalization) and a learning strategy guided by an empirical relation involving the distance, the inclination, and $\cal M$ for the signal-to-noise ratio (SNR) calculation. In fact, the use of an empirical SNR relation that prioritizes stronger signals early in training is a key aspect of the training strategy that allows the network to refine its weights more efficiently.

Additionally, it is important to note that the input data are pre-processed through whitening and adaptive normalization to improve resilience against noise fluctuations. These data consist of 4.25-second segments processed with a 1.4-second stride, with half of the segments containing injected signals placed randomly within 1.25-second intervals. Following whitening and edge trimming, the network receives 1-second inputs (per detector) sampled at 2 kHz.  Note, however, that while this configuration is effective for intermediate- and high-mass binaries, it is probably less suitable for low-chirp-mass systems, whose signals can extend over 20 seconds.

\section{\label{sec:Training_Data} Test Datasets}
\label{sec:test_datasets}

For this analysis, we used the MLGWSC-1 one-month dataset with real noise — defined by an offset of 0 and a noise and injection seed of 2514409456 — as a common reference across all test categories. In addition to this reference dataset, we generated 27 more one-month test datasets, resulting in a total of 28 datasets. These datasets differ in the following parameters: the slide buffer (which sets the maximum random time shift applied to the Livingston segments), the noise seed (which controls noise variation), the injection seed (which determines the set of injection parameters), and the start offset (which selects the starting point within the 81-day noise file provided in \cite{challenge1}). The 27 varied datasets are divided into three categories, each containing 9 datasets. With the shared reference dataset included in all three categories, each category contains 10 datasets in total. The goal is to independently assess the impact of noise, the effect of injections, and the combined influence of both. For reproducibility, all of these datasets were generated using the script provided for MLGWSC-1. 

\textit{Datasets with identical noise and varying injections.} In this category, all ten test sets share the same background noise, generated using a fixed noise seed (2514409456), a start offset of 0, and a slide buffer of 240 seconds. This configuration produces the exact same noise as used in the real-noise test of MLGWSC-1 \cite{challenge1}, as well as in \cite{AresGW_model,nagarajan2025identifyingmitigatingmachinelearning}. The only variation between these data sets comes from the injections, which were generated using different random seeds.

\textit{Datasets with varying noise and identical injections.} For this category, 10 injection files were created that contained identical injections (i.e., the same injection parameters) but assigned different GPS times. These files were generated by identifying which injections from the file with seed 2514409456 were included in the foreground set with real O3a noise in the MLGWSC-1 challenge \cite{challenge1}. However, each background dataset had a unique noise seed, and all, instead of the reference dataset which was created with a slide buffer of 240, feature a slide buffer of 900 seconds to accommodate larger shifts between the detector strains.

\textit{Datasets with both varying noise and injections.} 
These datasets were generated using varying noise offsets and different (but matching) noise and injection seeds\footnote{Each dataset uses matching noise and injection seeds; however, the noise and injection seeds are independent parameters. Matching values were chosen for simplicity in presentation.}, with the background datasets created using a slide buffer of 240 seconds.

The same three analyses will be performed twice: once on the data after removing only the GWTC-2 \cite{GWTC2} events, following the approach in \cite{challenge1,nagarajan2025identifyingmitigatingmachinelearning} , and once after removing all known events present in the O3a data from AresGW \cite{Koloniari2025}, GWTC-2.1 \cite{LIGOScientific:2021usb}, IAS \cite{IAS_O3a, IAS_higher_harmonics} and OGC \cite{OGC-3,OGC-4}  catalogs. This two-step approach enables a direct comparison with the analysis presented in \cite{nagarajan2025identifyingmitigatingmachinelearning}, suggesting that some of the conclusions drawn are affected by overlooked factors, such as dataset contamination.

For brevity, the results after total event removal for the categories "Datasets with varying noise and identical injections" and "Datasets with both varying noise and injections" are presented in Appendix~\ref{appendix:clean}.

\begin{figure*}[ht]
    \centering
    
    \includegraphics[width=0.31\textwidth]{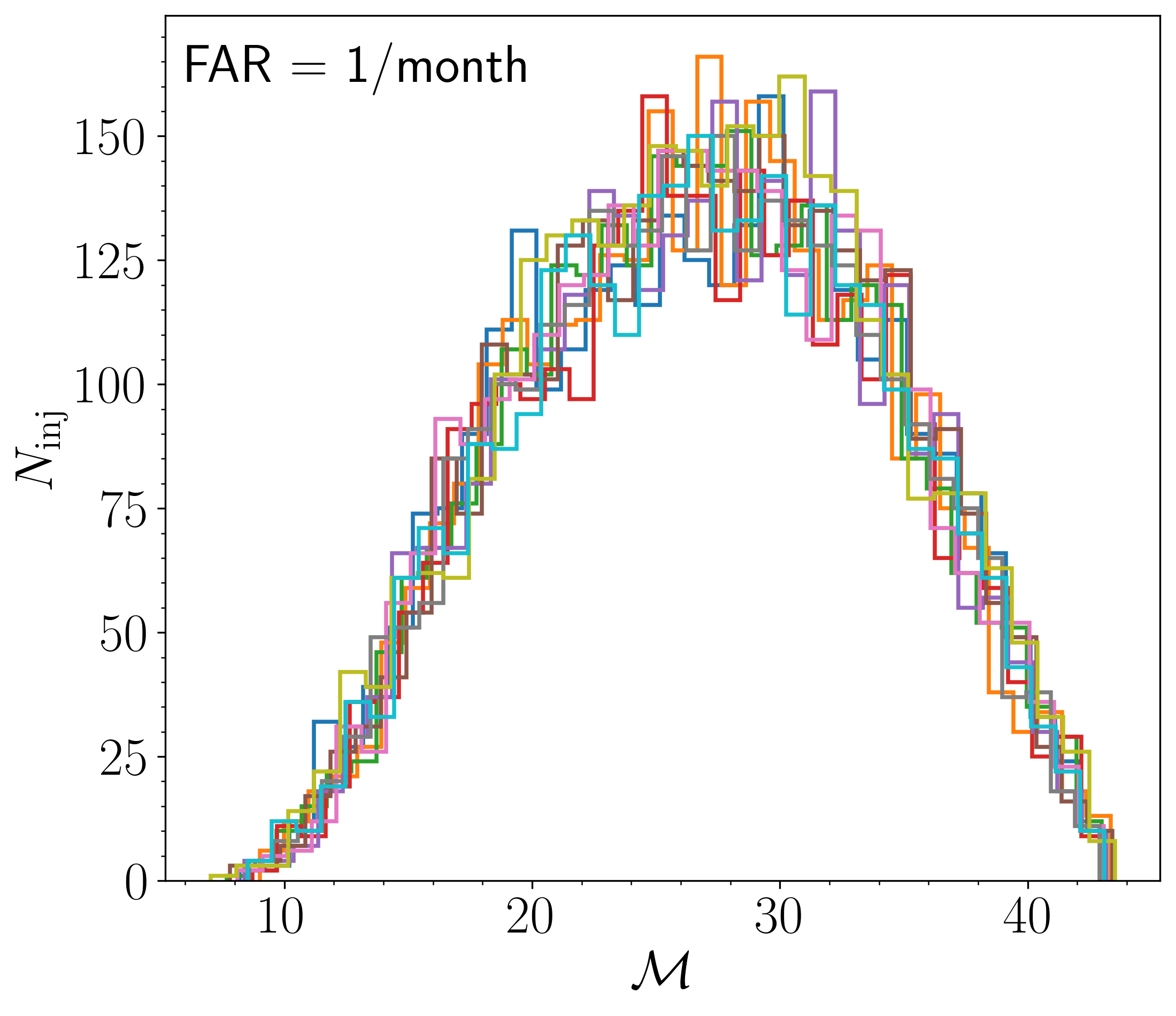} 
    \hspace{0.1cm} 
    \includegraphics[width=0.31\textwidth]{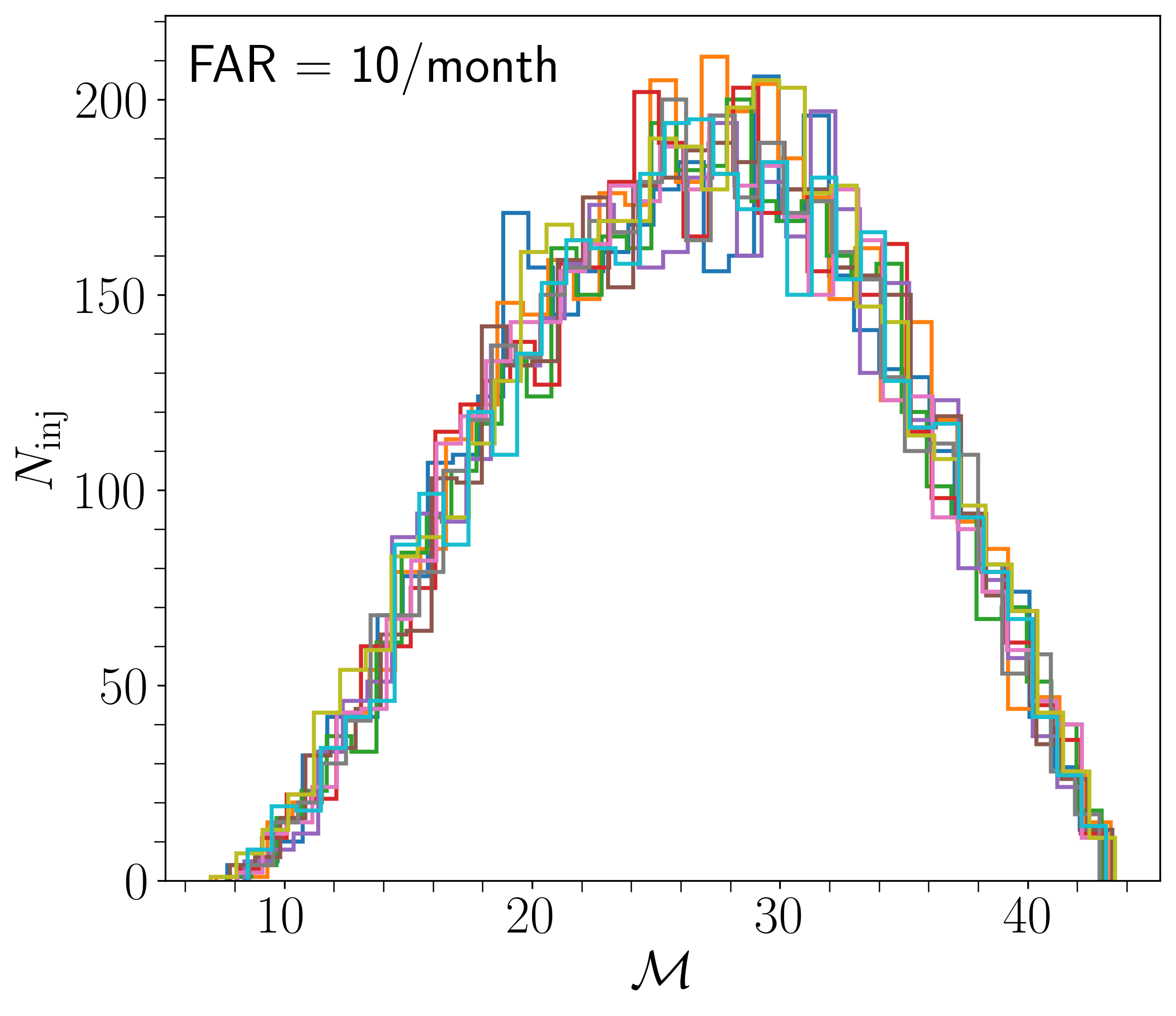} 
    \hspace{0.1cm} 
    \includegraphics[width=0.31\textwidth]{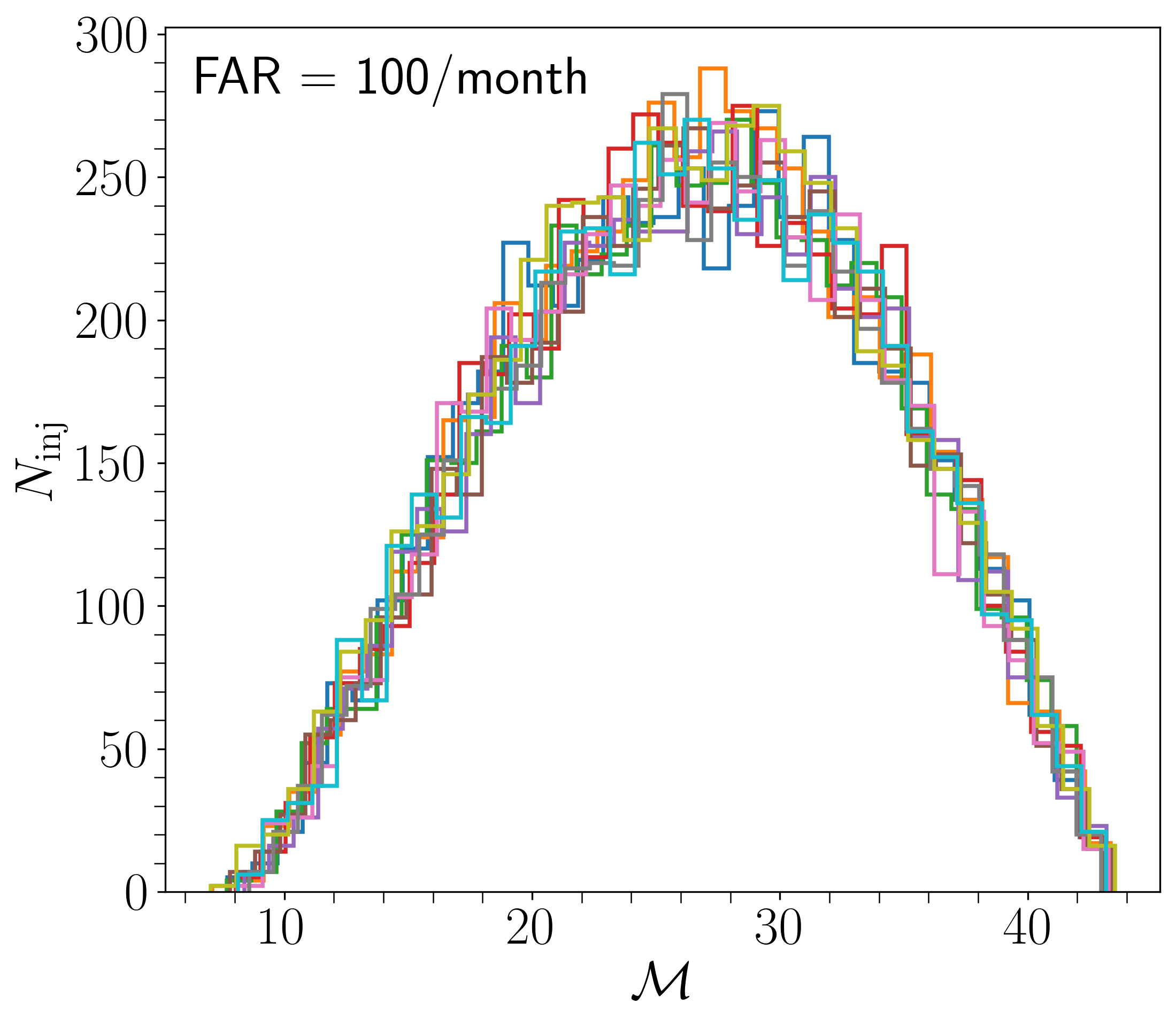}
    \vspace{0.5cm} 
    \includegraphics[width=0.7\textwidth]{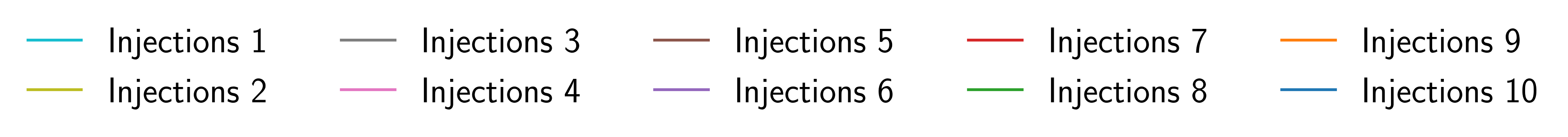}
    
    \caption{Number of detected injections as a function of chirp mass for various one-month datasets, using identical background noise (offset = 0 and noise seed = 2514409456), but with different signal injections. Each panel corresponds to a fixed false alarm rate threshold.}
    \label{fig:same_noise_diff_inj}
\end{figure*}

\section{\label{sec:performance}Performance Evaluation on Datasets with Event Contamination}

Here, we conduct three types of analysis on datasets that correspond to the scenarios described in Sec. \ref{sec:test_datasets}: (1) identical noise with varying signal injections, (2) varying noise with identical injections, and (3) both noise and injections varying. All three analyses in this section are performed using \textit{noise that still contains actual GW signals} not included in the GWTC-2 catalog \cite{GWTC2}. All included signals are listed in Table \ref{tab:appendix_events}, located in Appendix \ref{appendix_events}.

As mentioned in Sec.~\ref{sec:test_datasets}, the dataset with offset~0 and both the noise and injection seeds set to 2514409456 -- corresponding to real O3a detector noise from the two LIGO detectors, as used in~\cite{challenge1},~\cite{AresGW_model}, and~\cite{nagarajan2025identifyingmitigatingmachinelearning} -- serves as the reference dataset. It is included in all three dataset categories and is consistently depicted in light blue throughout all plots in this section.

Additionally, for a more comprehensive analysis, we computed the confidence intervals (CIs) for the mean, the standard deviation, and the coefficient of variation of two key metrics -- the number of detected injections and the sensitive distance --
across different FAR thresholds for each sample. In cases where the assumption of normality could not be rejected, we employed Student's $t$-distribution with 9 degrees of freedom (corresponding to a sample size of $n = 10$) to calculate the 95\% CI for the mean. Similarly, the chi-square distribution was used to derive the 95\% CI for the variance, from which the 95\% CI for the standard deviation was obtained by taking the square root of the bounds. Finally, the 95\% CI for the coefficient of variation is obtained by dividing the CI of the standard deviation by the sample mean. The CI for the mean will be presented in the form $\text{sample mean} \pm \text{margin of error}$, while the CIs for the standard deviation and the coefficient of variation will be presented as intervals $[\text{lower bound}, \text{upper bound}]$.

In cases where normality was rejected, we used the bias-corrected and accelerated (BCa) bootstrap method \cite{diciccio1996bootstrap} to estimate both the mean and the standard deviation, as it provides more accurate confidence intervals by adjusting for both bias and skewness in the sampling distribution. This approach is particularly suitable for small sample sizes or data with unknown or non-normal distributions, where standard bootstrap percentile intervals may be inadequate.

Moreover, note that we assessed normality in all the different samples using the Shapiro–Wilk test and Q--Q plots. However, for brevity, the Shapiro–Wilk p-values and Q--Q plots are presented only for cases where normality was rejected. A detailed description of the CI calculation methods and normalization tests used in all analyses can be found in Appendix~\ref{appendix:statistics}.

\subsection{Identical Noise and Varying Injections}
\label{sec:same_noise_diff_inj}

In this analysis, we utilize datasets constructed from a shared background file paired with distinct injection files, as described in Sec. \ref{sec:test_datasets}. The background used corresponds to a noise realization with seed 2514409456 and a start offset of 0, making it identical to the background file employed in the real-noise test of MLGWSC-1. Notably, although the offset is set to 0 and the duration spans an equivalent of 30 days, the end GPS time of the background file is 1253966059, which places it at the beginning of the third month of O3a. \textit{This implies that the background contains 11 additional confident GW events, including GW190511\_125545}, which was detected by AresGW model 2 with a very high ranking statistic \cite{Koloniari2025}. As we will see in Sec.~\ref{sec:performance_clean}, the inclusion of this particular event in the noise data plays a crucial role in the evaluation of AresGW model~1.

Table \ref{tab:AresGW_same_noise_diff_inj} reports the detected injection counts $N^{\rm F}$ across one-month datasets with varying injection configurations (different seeds) at the FAR thresholds of 1, 10, and 100  per month, along with total injections. It also provides CIs for the mean number of detected injections ($\mu_{\rm N}$), the standard deviation ($\sigma_{\rm N}$) and the coefficient of variation ($\sigma_{\rm N}/\bar\mu_{\rm N}$). Fig. \ref{fig:same_noise_diff_inj} complements this with distributions of chirp mass $\cal{M}$ for the same FAR thresholds.

Thus, for $\mathrm{FAR} = 1/\text{month}$, $\mu_{\rm N}$ is $2913 \pm 23$, with the standard deviation lying in the interval $\sigma_{\rm N} \in [22, 59]$. At $\mathrm{FAR} = 10/\text{month}$, $\mu_{\rm N}$ increases to $3855 \pm 29$, with $\sigma_{\rm N} \in [28, 75]$. Similarly, for $\mathrm{FAR} = 100/\text{month}$, $\mu_{\rm N}$ reaches $5384 \pm 42$, with $\sigma_{\rm N} \in [40, 107]$. Finally, we note that for this set of datasets, the performance variability for the number of detected injections remains low and consistent across all three FAR thresholds, the CIs for $\sigma_{\rm N}/\bar\mu_{\rm N}$ being $[0.4\%, 1.0\%]$, $[0.4\%, 1.0\%]$, and $[0.5\%, 1.2\%]$, respectively.

Now, let us perform the same analysis using the sensitive distance (see Appendix \ref{appendix:sensitive_distance}) at the same FAR thresholds as a metric. The goal is to assess whether one of the two metrics is more robust to variations using different one-month test datasets.

\begin{table}[t]
    \centering
    \caption{Detection performance evaluated on multiple one-month datasets containing identical real detector noise (offset = 0 and noise seed = 2514409456) but different sets of signal injections, using the number of detections $N^{\rm F}$ at fixed FAR as a metric. All datasets are still contaminated with a number of confident astrophysical signals not included in GWTC-2 \cite{GWTC2}.}
    \resizebox{0.44\textwidth}{!}{
    \begin{tabular}{|r|l|c|c|c|c|}
        \hline
        \rowcolor{blue!15} \# & Injection & $N^{\rm F}$ & $N^{\rm F}$  & $N^{\rm F}$ & Total \\
        \rowcolor{blue!15}  & Seed & ${\rm 1/month}$ & ${\rm10/month}$  & ${\rm 100/month}$ &  Injections \\
        \hline
        \rowcolor{blue!10}
        1 &2514409456  & 2892 & 3879 & 5436 &  95719\\
        2 & 12019 & 2949 & 3889 & 5415 & 95691\\
        \rowcolor{blue!10}
        3 & 10209 & 2924 & 3921 & 5429 & 95733\\
        4 &  9801  & 2923 & 3853 & 5375 & 95701\\
        \rowcolor{blue!10}
        5 &  6291  & 2935 & 3829 & 5261 & 95696\\
        6 & 555& 2904& 3824 & 5338 & 95698\\
        \rowcolor{blue!10}
        7 & 291& 2874& 3872 & 5459 & 95704\\
        8 &  93  & 2852 & 3778 & 5339 & 95711\\
        \rowcolor{blue!10}
        9 & 32 & 2949 & 3878 & 5386 & 95684 \\
        10 & 9 & 2927 & 3826 & 5406 & 95712 \\
        \hline
        \hline
        \rowcolor{blue!10}
        \multicolumn{2}{|c|}{$\mu_{\rm N}$ (mean)} & $2913 \pm 23$ &  $3855\pm 29$ & $5384 \pm 42$ & N/A \\
        \hline
        \rowcolor{blue!10}
        \multicolumn{2}{|c|}{$\sigma_{\rm N}$ (std. dev.)}  & $\in[22, 59]$  & $\in[28, 75]$  & $\in[40, 107]$  & N/A \\
        \hline
        \rowcolor{blue!10}
        \multicolumn{2}{|c|}{$\sigma_{\rm N}/\bar\mu_{\rm N}$ }  & $\in[0.8\%, 2.0\%]$  & $\in[0.7\%, 1.9\%]$ & $\in[0.7\%, 2.0\%]$  &  N/A \\
        \hline
    \end{tabular}
}
    \label{tab:AresGW_same_noise_diff_inj}
\end{table}

\begin{figure}[t]
  \centering
  \includegraphics[width=0.92\linewidth]{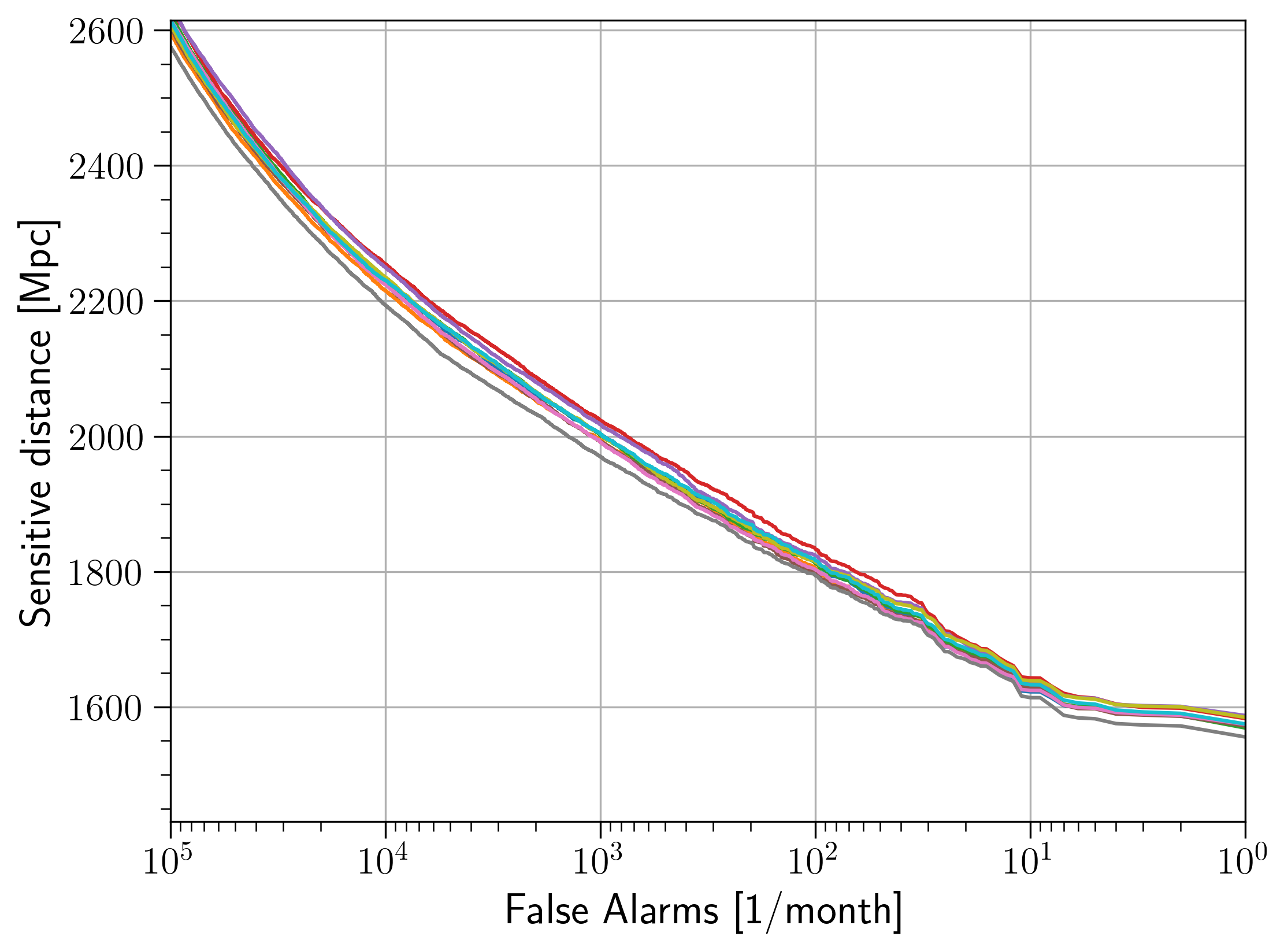}
  \caption{Sensitive distance vs. FAR, using identical background noise (same as in \cite{challenge1,nagarajan2025identifyingmitigatingmachinelearning}) with different signal injections. The color scheme is as in Fig. \ref{fig:same_noise_diff_inj}.}  \label{fig:sensitivity_same_noise_diff_inj}
\end{figure}

\begin{table}[t]
    \centering
    \renewcommand{\arraystretch}{1.1} 
    \caption{Sensitive distances achieved at various FAR thresholds, evaluated across multiple one-month datasets with identical real detector noise (offset = 0 and noise seed = 2514409456) but different sets of signal injections.}
    \footnotesize
    \resizebox{0.39\textwidth}{!}{
    \begin{tabular}{|r|c|c|c|c|}
        \hline
        \rowcolor{blue!15} \# & Injection & $S$ (Mpc) & $S$ (Mpc) & $S$ (Mpc)\\
        \rowcolor{blue!15}  & Seed & 1/month & 10/month  & 100/month \\
        
        \hline
        \rowcolor{blue!10}
        1 & 2514409456  & 1574.94 & 1632.87 & 1816.48\\
        2 & 12019 & 1584.71 & 1638.79 & 1814.27\\
        \rowcolor{blue!10}
        3 & 10209 & 1555.73 & 1613.79 & 1793.98\\
        4 & 9801 & 1574.38 & 1624.49 & 1802.85\\
        \rowcolor{blue!10}
        5 & 6291 & 1571.94 & 1628.12 & 1797.46\\
        6 & 555 & 1587.36 & 1637.09 & 1823.07\\
        \rowcolor{blue!10}
        7 & 291 & 1582.93 & 1642.93 & 1833.30\\
        8 & 93 & 1569.01 & 1624.76 & 1812.07\\
        \rowcolor{blue!10}
        9 & 32 &  1571.75 &  1626.39 & 1804.75\\
        10 & 9 &  1573.24 &  1622.46 & 1814.49\\
        \hline
        \hline
        \rowcolor{blue!10}
        \multicolumn{2}{|c|}{$\mu_{\rm S}$ (mean)} & $1574.6 \pm 6.5$ & $1629.2 \pm 6.3$& $1811.3 \pm 8.5$ \\
        \hline
        \rowcolor{blue!10}
        \multicolumn{2}{|c|}{$\sigma_{\rm S}$ (std. dev.)}  &  $ \in [6.2, 16.5]$  & $ \in[6.0, 16.0]$  &  $ \in[8.2, 21.7]$  \\
        \hline
        \rowcolor{blue!10}
        \multicolumn{2}{|c|}{$\sigma_{\rm S}/\bar\mu_{\rm S}$ }  & $\in[0.4\%, 1.0\%]$ & $\in[0.4\%, 1.0\%]$ & $\in[0.5\%, 1.2\%]$  \\
        \hline
    \end{tabular}
    }  \label{tab:sensitivity_distance_same_noise_diff_inj}
\end{table}

The sensitive distance $S$ as a function of FAR is shown in Fig.~\ref{fig:sensitivity_same_noise_diff_inj}. The corresponding sensitive distances at FAR thresholds of 1, 10, and 100 per month, along with the CIs for the sample mean ($\bar\mu_{\rm S}$), standard deviation ($\sigma_{\rm S}$) and coefficient of variation ($\sigma_{\rm S}/\bar\mu_{\rm S}$), are summarized in Table~\ref{tab:sensitivity_distance_same_noise_diff_inj}.

For FAR = 1/month, $\mu_{\rm S}$ is $1574.6 \pm 6.5$~Mpc, with $\sigma_{\rm S} \in [6.2, 16.5]$~Mpc. When the FAR increases to 10/month, $\mu_{\rm S}$ increase to $1629.2 \pm 6.3$~Mpc, with $\sigma_{\rm S} \in [6.0, 16.0]$~Mpc. At the highest tested FAR of 100/month, $\mu_{\rm S}$ increases further to $1811.3 \pm 8.5$~Mpc, with the confidence interval for $\sigma_{\rm S}$ given by $[8.2, 21.7]$~Mpc. Furthermore, again, as in the case of the number of detected injections, the performance in the different datasets using the sensitive distance as a metric is robust, with CIs for $\sigma_{\rm S}/\bar\mu_{\rm S}$ equal to $[0.4\%, 1.0\%]$, $[0.4\%, 1.0\%]$, and $[0.5\%, 1.2\%]$, at FAR = 1, 10, and 100 per month, respectively.

The relatively low variance in detection rate -- in all three different FAR thresholds and for both metrics -- across the test sets that contained identical background noise but varied signal injections,  drawn from the same parameter distributions, is notable. This suggests that the network's performance is robust to the specific realizations of the injected signals. This implies that the model generalizes well across different instances of the target signal distribution and is not overly sensitive to particular waveform characteristics or random variations within that distribution. Moreover, since the noise was held constant, the consistency in detection rates further indicates that the model's performance is not being skewed by overfitting to particular signal-noise interactions, but rather reflects a stable and reliable sensitive to the signal class as a whole.

Additionally, these results anticipate that significant performance variability will be noise-dominated: As we just mentioned, the consistency in detection performance across different signal realizations, with fixed noise, indicates that the model's sensitivity is largely invariant to specific signal parameters drawn from the same distribution. This suggests that, in general, variability in detection performance is more likely to stem from changes in the noise environment than from differences in the injected signals -- a point that will be demonstrated further in the following two subsections. Therefore, the noise characteristics play a more dominant role in shaping the overall detection effectiveness.

\subsection{Varying Noise and Identical Injections}

\begin{figure*}[t]
    \centering
    \includegraphics[width=0.31\textwidth]{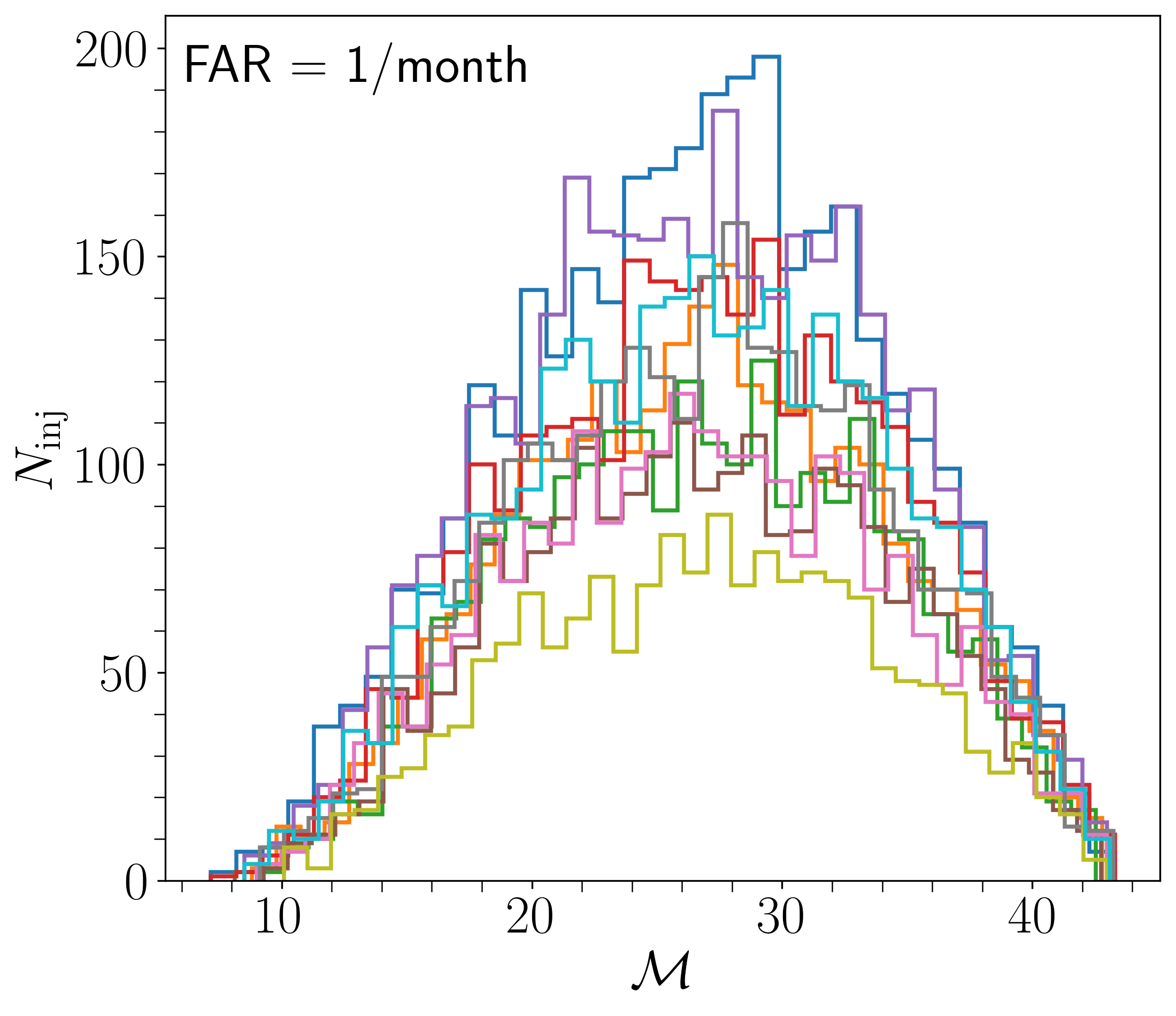} 
    \hspace{0.1cm} 
    \includegraphics[width=0.31\textwidth]{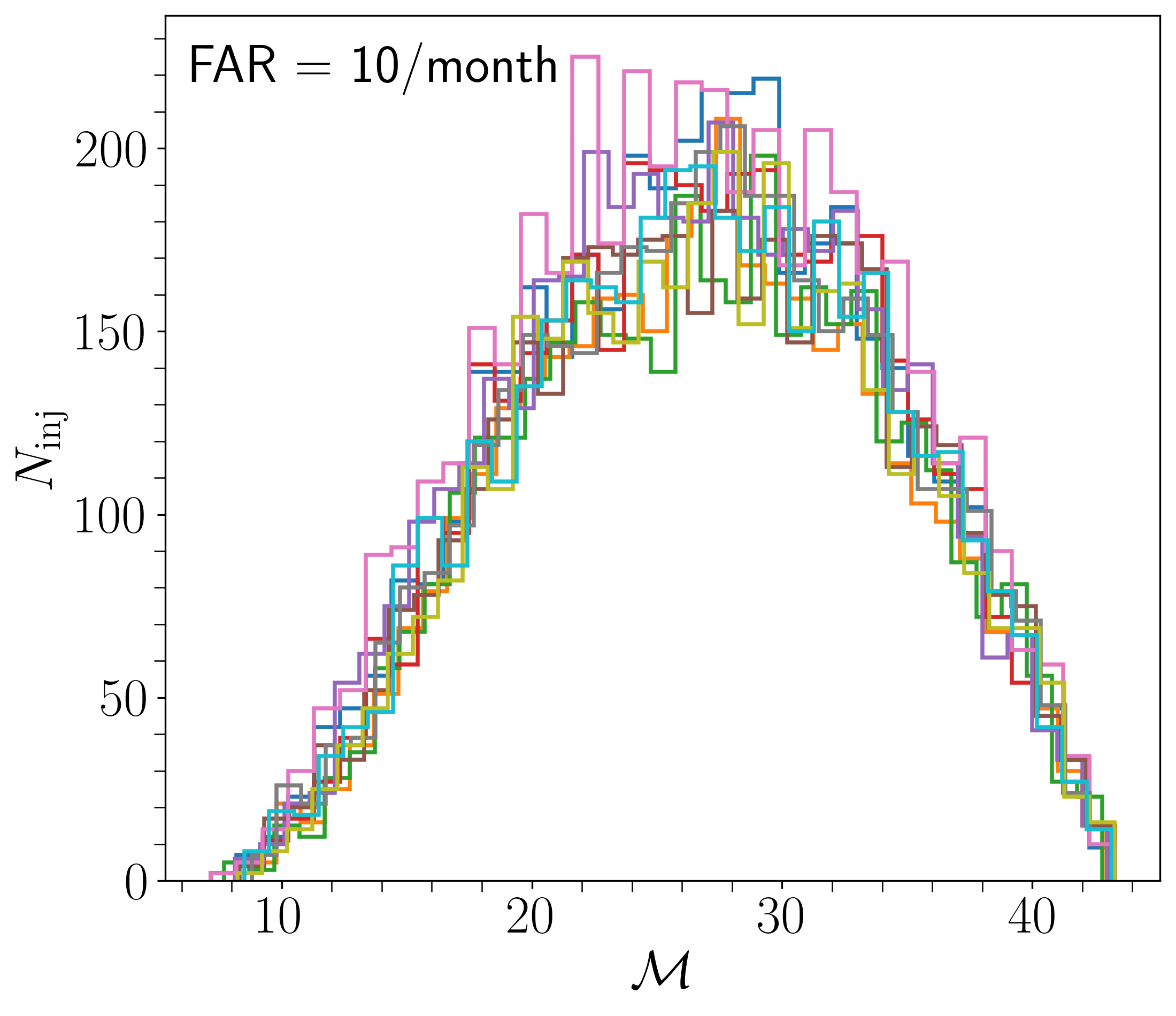} 
    \hspace{0.1cm} 
    \includegraphics[width=0.31\textwidth]{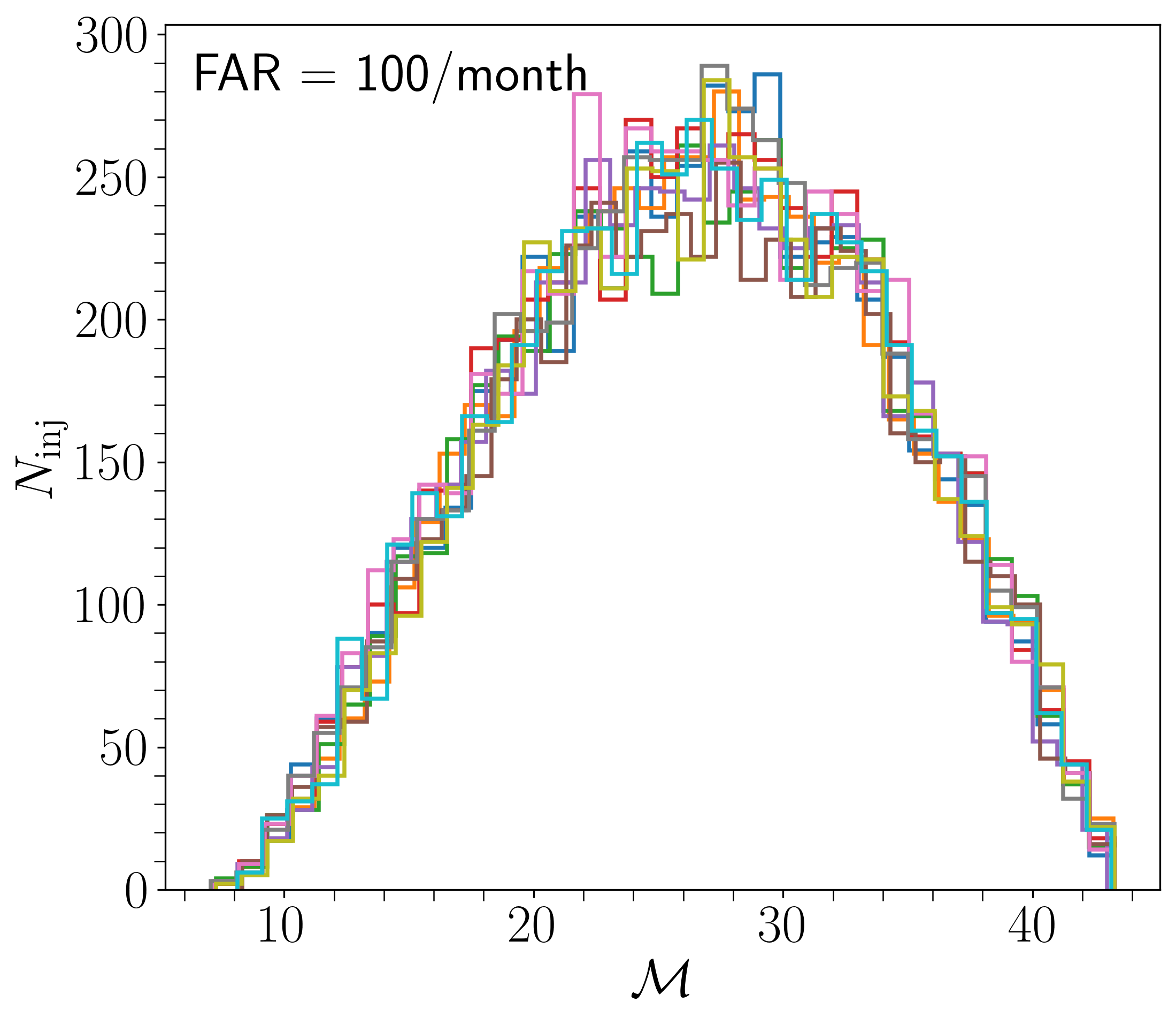}
    \vspace{0.5cm} 
    \includegraphics[width=0.6\textwidth]{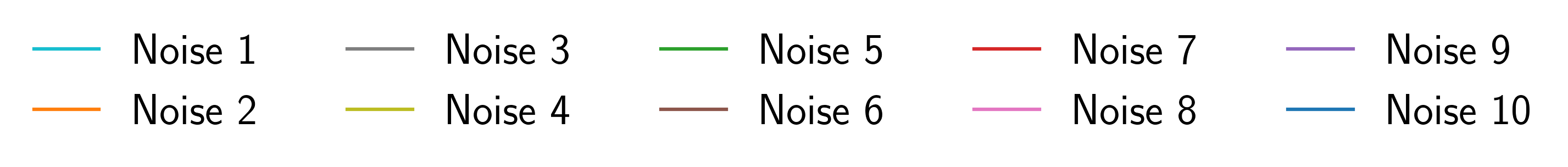}
    \caption{As in Fig \ref{fig:same_noise_diff_inj}, but using different real detector noise and a fixed set of simulated signal injections.}
    \label{fig:diff_noise_same_inj}
\end{figure*}

\label{sec:diff_noise_same_inj}
The datasets used in this analysis were generated using 10 injection files, each containing the same injections as those in the MLGWSC-1 real-noise dataset, but with different GPS times. Specifically, we started with the foreground file of the MLGWSC-1 real-noise dataset, identified which injections from the MLGWSC-1 injection set (generated with seed 2514409456) were included, and then adjusted their GPS times to align with the different background noise files. Thus, in this set of datasets, at least 99.9\% of the injections contained in the foreground files are common across all test sets, with minor variations in the number of injections due to padding and noise gaps.  

Note that although all source-related parameters — including the sky location — remain identical across the different datasets, the injections are performed at different GPS times, leading to variations in the projected waveforms, due to Earth's rotation. Consequently, the datasets differ not only in terms of noise realizations but also in the detectors’ responses to the signals.

Regarding the contamination of the noise, it is important to point out that, \textit{these datasets, combined, still contain approximately 40 real signals}.

\begin{table}[t]
    \centering
    \renewcommand{\arraystretch}{1.1} 
    \caption{As in Table \ref{tab:AresGW_same_noise_diff_inj}, but using different real detector noise and a fixed set of simulated signal injections.}
    \renewcommand{\arraystretch}{1.1} 
    \footnotesize
    \resizebox{0.5\textwidth}{!}{
    \begin{tabular}{|r|l|c|c|c|c|c|}
        \hline
        \rowcolor{blue!15} \# & Offset & Noise & $N^{\rm F}$ & $N^{\rm F}$  & $N^{\rm F}$ & Total \\
        \rowcolor{blue!15} &(days)& Seed & ${\rm 1/month}$ & ${\rm10/month}$  & ${\rm 100/month}$ &  Injections \\
        \hline
        \rowcolor{blue!10}
        1 & 0 & 2514409456 & 2892 & 3879 & 5436 & 95719 \\
        2 & 0 & 500724 & 2597 & 3652 & 5372 & 95697 \\
        \rowcolor{blue!10}
        3 & 0 & 16 & 2733 & 3924 & 5346 & 95730\\
        4 & 10 & 145 & 1668 & 3661 & 5167 & 95702\\
        \rowcolor{blue!10}
        5 & 20 & 3199 & 2307 & 3565 & 5219 & 95694\\
        6 & 20 & 313 & 2191 & 3878 & 5181 & 95729\\
        \rowcolor{blue!10}
        7 & 30 & 1009 & 2788 & 3851 & 5387 & 95702\\
        8 & 30 & 2 & 2242 & 4361 & 5462 & 95710\\
        \rowcolor{blue!10}
        9 & 40 & 6 & 3470 & 4051 & 5371 & 95693\\
        10 & $\approx$46.3 & 7897 & 3456 & 3949 & 5274 & 95725\\
        \hline
        \hline
        \rowcolor{blue!10}
        \multicolumn{3}{|c|}{$\mu_{\rm N}$ (mean)} & $2634 \pm 404 $ & $ 3871\pm163 $ & $ 5322\pm 76 $ &  N/A \\
        \hline
        \rowcolor{blue!10}
        \multicolumn{3}{|c|}{$\sigma_{\rm N}$ (std. dev.)}  & $\in[388, 1030]$  & $\in[157, 417]$  & $\in[72, 192]$  &  N/A \\
        \hline
        \rowcolor{blue!10}
        \multicolumn{3}{|c|}{$\sigma_{\rm N}/\bar\mu_{\rm N}$ }  & $\in[14.7\%, 39.1\%]$  & $\in[4\%, 10.7\%]$  & $\in[1.4\%, 3.6\%]$  &  N/A \\
        \hline
    \end{tabular}
    }
    \label{tab:AresGW_diff_noise_same_inj}
\end{table}

\begin{figure}[t]
  \centering
  \includegraphics[width=0.92\linewidth]{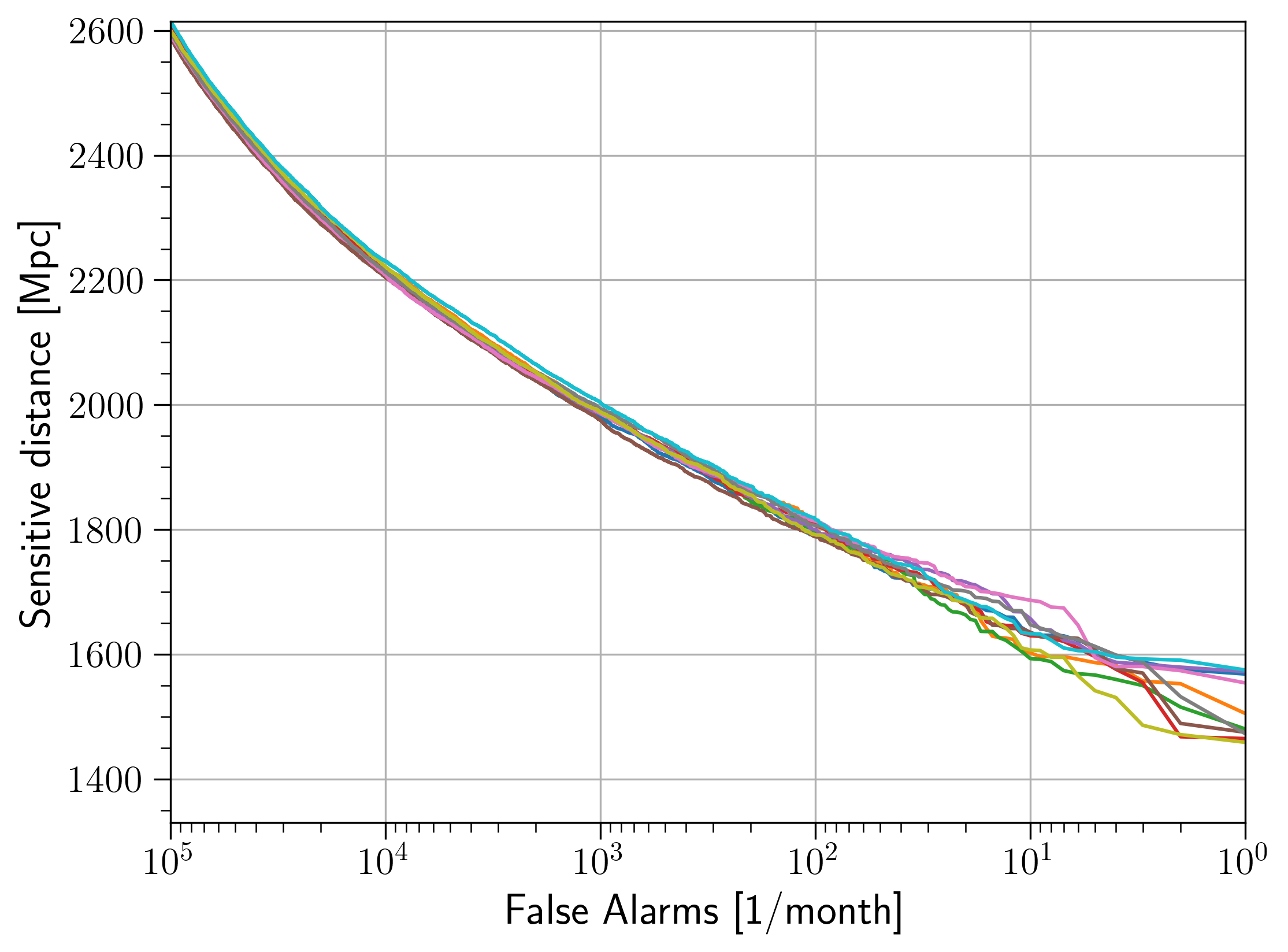}
  \caption{As in Fig. \ref{fig:sensitivity_same_noise_diff_inj} but using different real detector noise and a fixed set of simulated signal injections. The color scheme is as in Fig. \ref{fig:diff_noise_same_inj}.}  \label{fig:sensitivity_diff_noise_same_inj}
\end{figure}

\begin{figure}[t]
  \centering
  \includegraphics[width=0.82\linewidth]{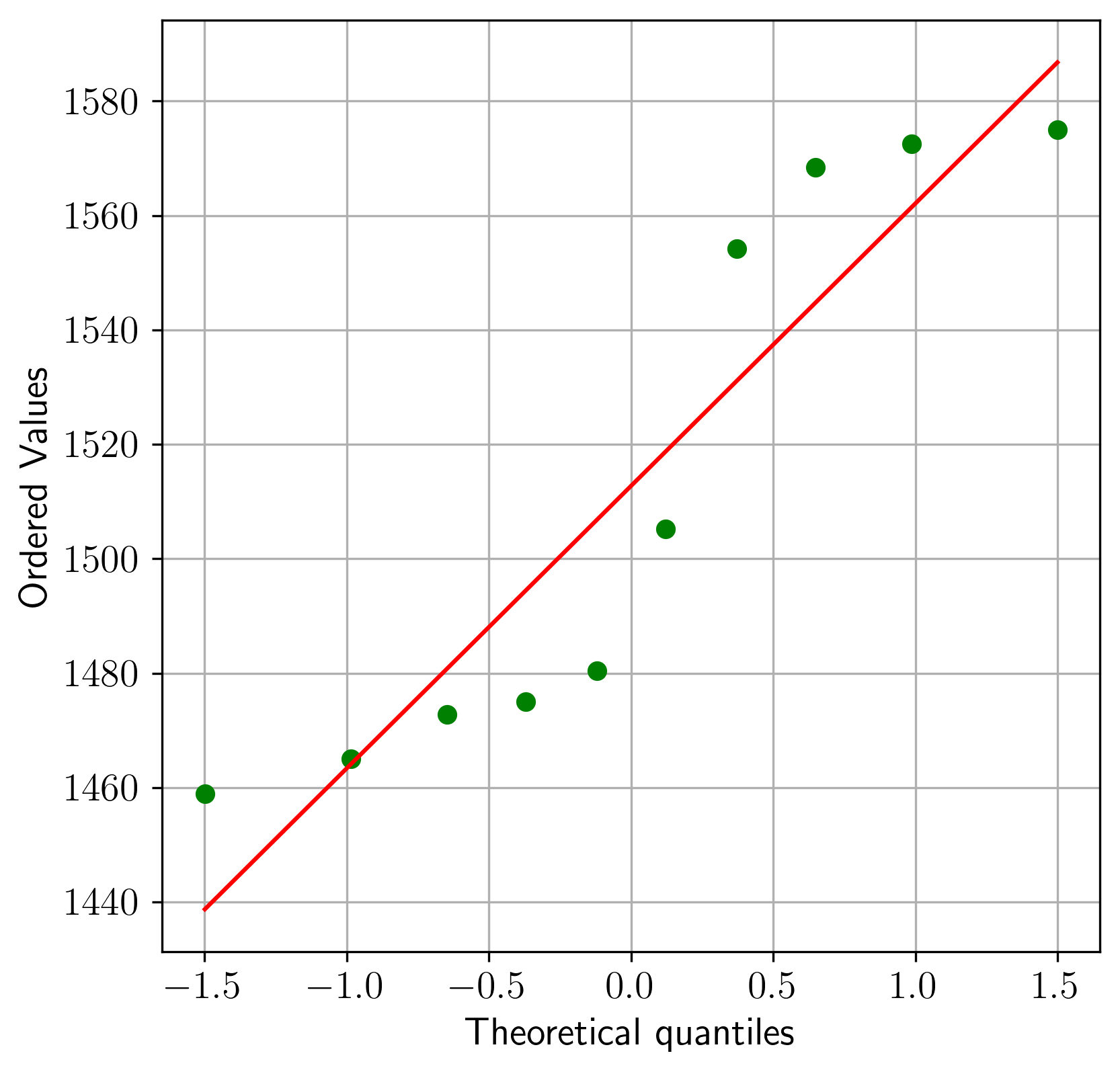}
  \caption{Q-Q plot of the sample data for the sensitive distance at FAR = 1/month.} 
  \label{fig:qq-sensitive_distance_same_injections}
\end{figure}

\begin{table}[t]
    \centering
    \renewcommand{\arraystretch}{1.1} 
    \caption{As in Table \ref{tab:sensitivity_distance_same_noise_diff_inj}, but using different real detector noise and a fixed set of simulated signal injections.}
    \footnotesize
    \resizebox{0.45\textwidth}{!}{
    \begin{tabular}{|r|l|c|c|c|c|}
        \hline
        \rowcolor{blue!15} \# & Offset  & Noise & $S\text{ (Mpc)}$ & $S\text{ (Mpc)}$& $S\text{ (Mpc)}$ \\
        \rowcolor{blue!15} &  (days) & Seed & 1/month & 10/month  & 100/month \\
        \hline
        \rowcolor{blue!10}
        1 & 0 & 2514409456 & 1574.94 & 1632.87 & 1816.48\\
        2 & 0 & 16 & 1472.77 & 1647.46 & 1806.78\\
        \rowcolor{blue!10}
        3 & 0 & 500724 & 1505.19 & 1602.31 & 1808.43\\
        4 & 10 & 145 & 1458.94 & 1606.98 & 1791.65\\
        \rowcolor{blue!10}
        5 & 20 & 3199 & 1480.37 & 1593.17 & 1796.74\\
        6 & 20 & 313 & 1475.02 & 1635.43 & 1788.59\\
        \rowcolor{blue!10}
        7 & 30 & 1009 & 1464.98 & 1629.82 & 1810.94\\
        8 & 30 & 2 & 1554.19 & 1686.71 & 1813.51\\
        \rowcolor{blue!10}
        9 & 40 & 6 & 1572.54 & 1657.21 & 1799.59\\
        10 & $\approx$ 46.3 & 7897 & 1568.46 & 1634.82 & 1790.24\\
        \hline
        \hline
        \rowcolor{blue!10}
        \multicolumn{3}{|c|}{$\mu_{\rm S}$ (mean)} & $ \in [1486.4, 1543.1]$ & $1632.7\pm19.8$ & $1802.3\pm7.3$ \\
        \hline
        \rowcolor{blue!10}
        \multicolumn{3}{|c|}{$\sigma_{\rm S}$ (std. dev.)}  &  $ \in [40.5, 56.3]$  & $ \in[19.1, 50.6]$  &  $ \in[7.0, 18.7]$  \\
        \hline
        \rowcolor{blue!10}
        \multicolumn{3}{|c|}{$\sigma_{\rm S}/\bar\mu_{\rm s}$ }  & $\in[2.7\%, 3.7\%]$  & $\in[1.2\%, 3.1\%]$  & $\in[0.4\%, 1.0\%]$  \\
        \hline
    \end{tabular}
    }
    \label{tab:sensitivity_diff_noise_same_inj}
\end{table}

\begin{figure*}[t]
    \centering
    \includegraphics[width=0.31\textwidth]{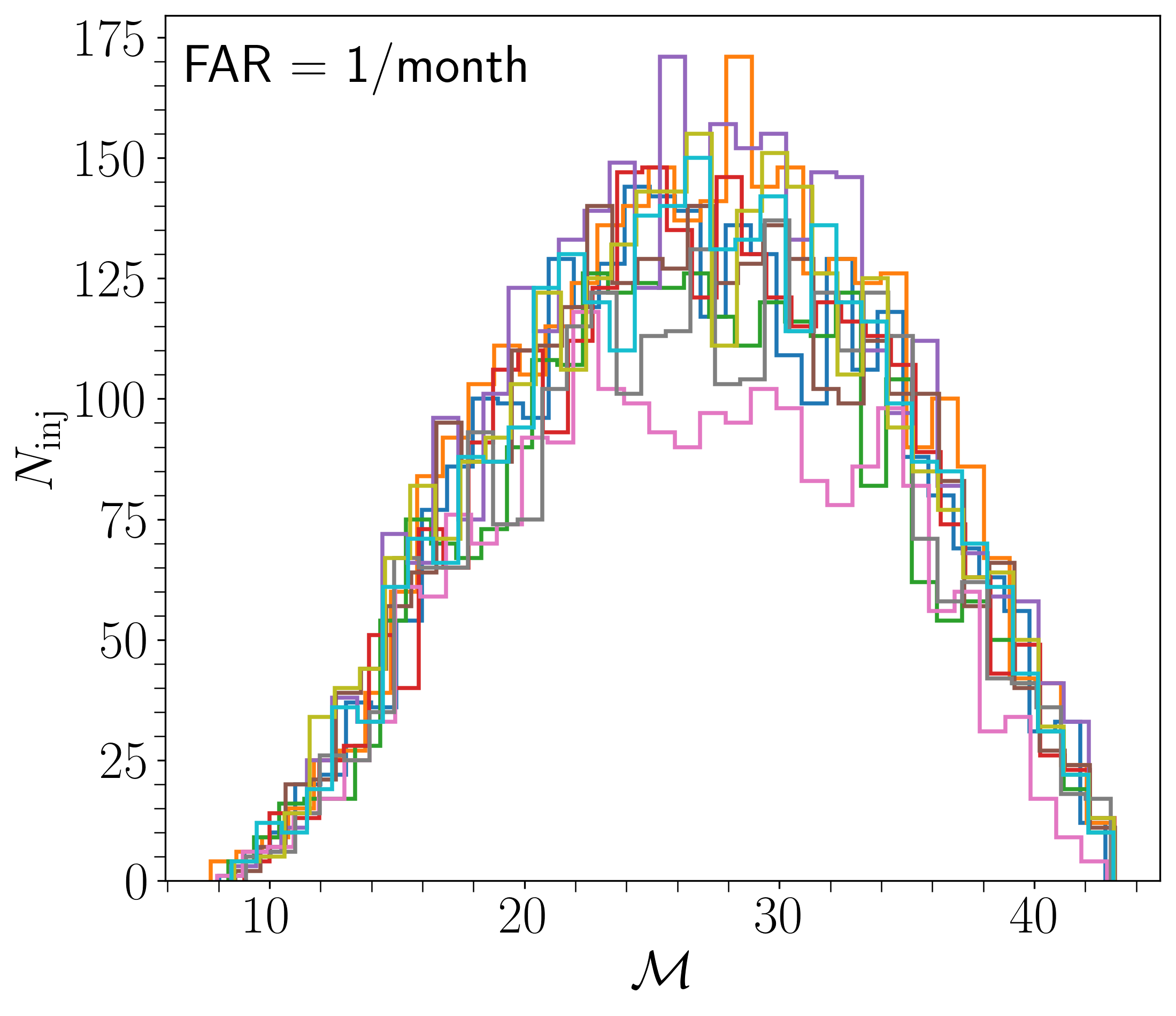} 
    \hspace{0.1cm} 
    \includegraphics[width=0.31\textwidth]{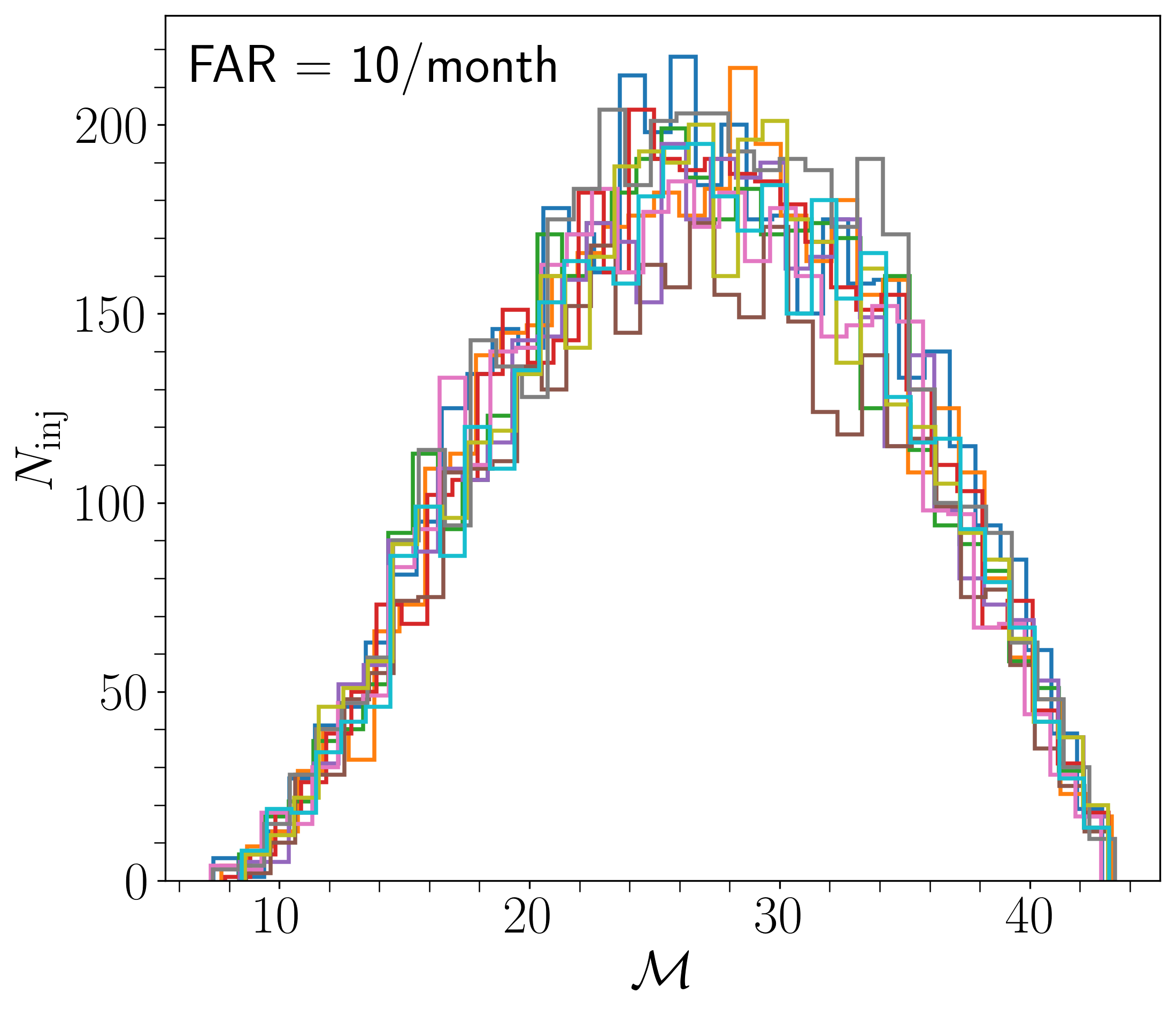} 
    \hspace{0.1cm} 
    \includegraphics[width=0.31\textwidth]{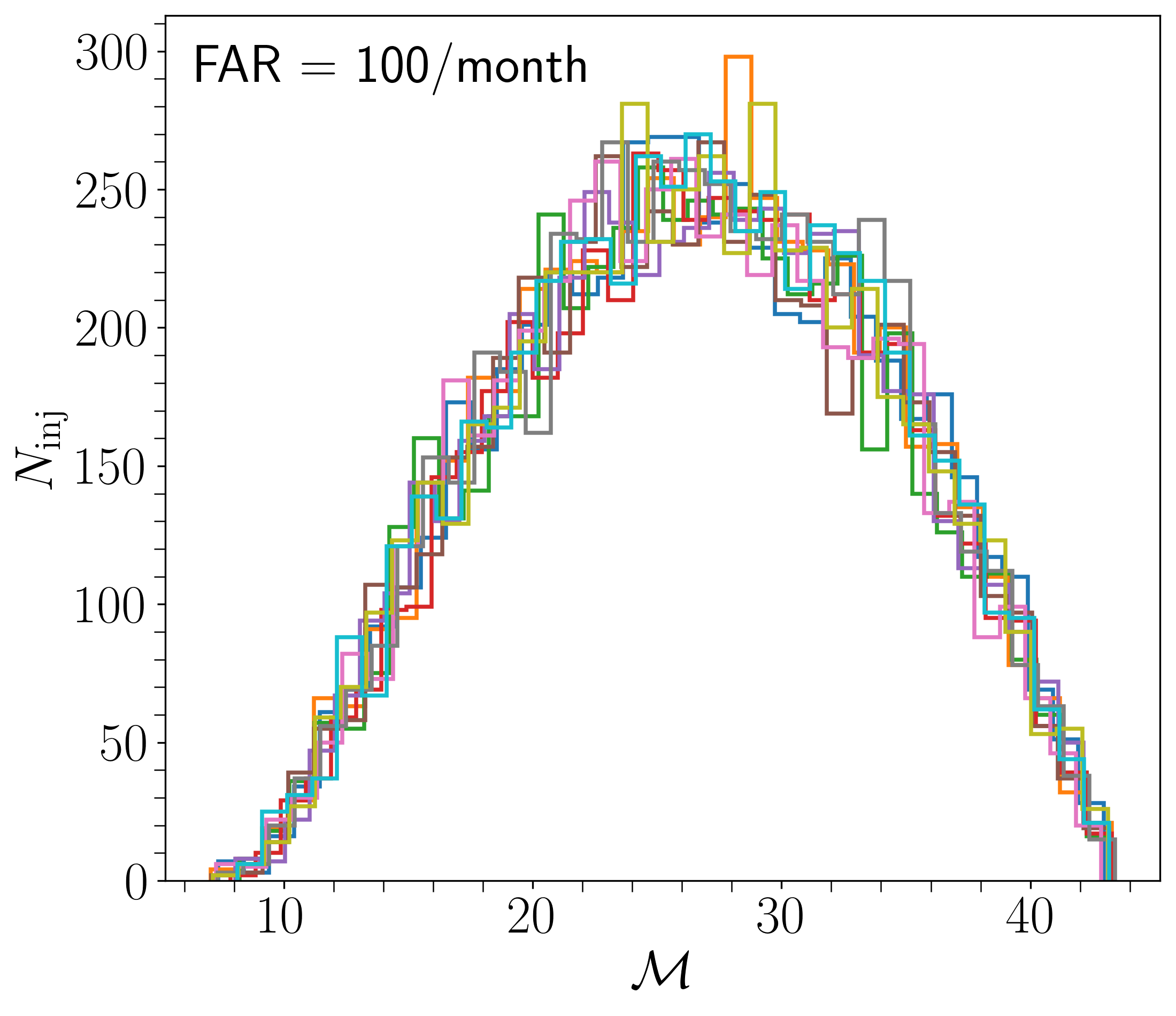}
    \includegraphics[width=0.9\textwidth]{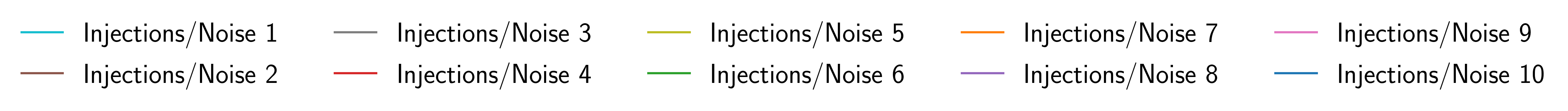}
    \caption{As in Fig \ref{fig:same_noise_diff_inj}, but using various instances of both real detector noise and simulated signal injections.}
    \label{fig:diff_noise_diff_inj}
\end{figure*}

Moving on to the analysis, similarly to Sec. \ref{sec:same_noise_diff_inj}, table \ref{tab:AresGW_diff_noise_same_inj} summarizes the number of detected injections in one-month datasets with varying noise realizations (different seeds and offsets) across different FAR thresholds, while also including the total number of injections present in each dataset and the CIs for $\mu_{\rm N}$, $\sigma_{\rm N}$ and $\sigma_{\rm N}/\bar\mu_{\rm N}$. Furthermore, Fig. \ref{fig:diff_noise_same_inj} shows histograms of detected injections as a function of $\cal{M}$ at fixed FAR thresholds. 

Importantly, the substantial variance observed in Fig. \ref{fig:diff_noise_same_inj} at FAR = 1/month contrasts sharply with the results in Fig.~\ref{fig:same_noise_diff_inj}, where identical noise and varying injections yielded much more consistent performance. This highlights the dominant role of noise fluctuations in driving performance variability under stricter detection thresholds, even when the injected signals remain fixed and differ only due to antenna response in addition to noise variations. We can attribute these results to noise fluctuations rather than antenna responses, given the consistent performance observed in the case of test sets presented in Sec. \ref{sec:same_noise_diff_inj}.

In particular, for FAR = 1/month, $\mu_{\rm N}$ was $2634 \pm 404$, while $\sigma_{\rm N} \in [388, 1030]$. For FAR = 10/month, $\mu_{\rm N}$ was $3871\pm 163$, and the CI for $\sigma_{\rm N}$ was $[157, 417]$. Similarly, for FAR = 100/month, $\mu_{\rm N}$ was $5322\pm 76$, with $\sigma_{\rm N} \in [72, 192]$. Finally, the CIs for $\sigma_{\rm N}/\bar\mu_{\rm N}$ at the FAR thresholds of 1/month, 10/month, and 100/month are $[14.7\%, 39.1\%]$, $[4\%, 10.7\%]$, and $[1.4\%, 3.6\%]$, respectively.

Note that for FAR = 1/month, the CIs for $\mu_{\rm N}$ and $\sigma_N$ are particularly wide, consistent with the expectations formed from Fig.~\ref{fig:diff_noise_same_inj}. In fact, \textit{the upper bound of the CI for $\sigma_{\rm N}/\bar{\mu}_{\rm N}$ in the case of FAR = 1/month is 39.1\%}. Similarly, for a FAR of 10/month, the CI for $\sigma_{\rm N}$ remains fairly wide, further indicating performance variability. The situation is somewhat improved at FAR = 100/month, where the CIs are narrower, suggesting that a single measurement may be more representative of the neural network's efficiency. However, a FAR of 100/month corresponds to 1,200/year, which is far too high to support a credible detection claim.

Moving on to the sensitive distance analysis, the results for all datasets are presented in Table \ref{tab:sensitivity_diff_noise_same_inj} and Fig. \ref{fig:sensitivity_diff_noise_same_inj}. 
For this metric of sensitivity, the increase in variance was much smaller. In fact, at FAR = 1/month, the sample mean for the sensitive distance ($\bar\mu_{\rm S}$) was 1512.74 Mpc with the CI being $[1486.4, 1543.1]$ Mpc. The corresponding CI for $\sigma_{\rm S}$ was equal to $[40.5, 56.3]$ Mpc. At FAR = 10/month, $\mu_{\rm S}$ increases to $1632.7\pm19.8$ Mpc and $\sigma_{\rm S} \in [19.1, 50.6]$ Mpc. Finally, at FAR = 100/month, $\mu_{\rm S}$ reaches $1802.3\pm7.3$ Mpc. The CI for $\sigma_{\rm S}$ tightens to $[7.0, 18.7]$ Mpc. When it comes to $\sigma_{\rm S}/\bar\mu_{\rm S}$, the CIs for FAR = 1/month, 10/month, and 100/month are $[2.7\%, 3.7\%]$ Mpc, $[1.2\%, 3.1\%]$ Mpc, and $[0.4\%, 1.0\%]$ Mpc, respectively.

Note that for FAR = 1/month, the data were not normally distributed. The p-value of the Shapiro--Wilk test was 0.0269; thus, a non-parametric bootstrap was implemented to compute the CI. The Q-Q plot (Fig.~\ref{fig:qq-sensitive_distance_same_injections}) supports this decision, showing a gentle S-shaped pattern — a visual indication of lighter tails than a normal distribution, meaning fewer extreme values than expected under normality. This pattern may also reflect slight skewness, although a classic S-shape is more commonly associated with differences in kurtosis (tailedness). However, it is important to note that the Q–Q plot does not suggest a strong departure from normality overall.

Thus, we conclude that \textit{the sensitive distance is more robust than the number of detected injections, particularly at lower FAR values}. For instance, at FAR = 1/month, the upper bound of the CI for $\sigma/\bar\mu$ of the sensitive distance corresponds to approximately 3.6\%, compared to around 39\% for the number of detected injections. Additionally, the upper bound of the CI for $\sigma_{\rm S}/\bar\mu_{\rm S}$ at FAR = 1/month is equal to $\sigma_{\rm N}/\bar\mu_{\rm N}$ at FAR = 100/month.

\subsection{\label{sec:diff_noise_diff_inj}  Both Varying Noise and Injections}

\begin{table}[t]
    \centering
    \renewcommand{\arraystretch}{1.1} 
    \caption{As in Table \ref{tab:AresGW_same_noise_diff_inj}, but using various instances of both real detector noise and simulated signal injections.}
    \footnotesize
    \resizebox{0.5\textwidth}{!}{
    \begin{tabular}{|r|l|c|c|c|c|c|}
        \hline
        \rowcolor{blue!15} \# & Offset  & Injection/ & $N^{\rm F}$ & $N^{\rm F}$  & $N^{\rm F}$ & Total \\
        \rowcolor{blue!15} & (days) &Noise Seed & 1/month & 10/month  & 100/month & Injections \\
        \hline
        \rowcolor{blue!10}
        1 & 0 & 2514409456 &2892  &3879  &5436  &95719 \\
        2 & 0 & 4 &2863  &3492  &5092  &95698 \\
        \rowcolor{blue!10}
        3 & 5 & 2371 &2618  &4122  &5294  &95687 \\
        4 & 15 & 12 &2847  &3933  &5154  &95737 \\
        \rowcolor{blue!10}
        5 & 25 & 7419 &2970  &3979  &5229  &95680 \\
        6 & 27 & 1213 &2566  &3959  &5193  &95704 \\
        \rowcolor{blue!10}
        7 & 30 & 102 &3051  &3947  &5238  &95688 \\
        8 & 35 & 1823 &3181  &3888  &5291  &95703 \\
        \rowcolor{blue!10}
        9 & 41 & 601 &2158  &3773  &5221  &95697 \\
        10 & 45 & 1000000 &2821  &4121  &5289  &95684 \\
        \hline
        \hline
        \rowcolor{blue!10}
        \multicolumn{3}{|c|}{$\mu_{\rm N}$ (mean)} & $2797\pm 207$ & $3909\pm129 $ & $5244\pm67 $ &   N/A \\
        \hline
        \rowcolor{blue!10}
        \multicolumn{3}{|c|}{$\sigma_{\rm N}$ (std. dev.)}  & $\in[199, 528]$  & $\in[124, 330]$  & $\in[64, 170]$  & N/A \\
        \hline
        \rowcolor{blue!10}
        \multicolumn{3}{|c|}{$\sigma_{\rm N}/\bar\mu_{\rm N}$ }  & $\in[7.1\%,18.9\%]$  & $\in[3.2\%, 8.4\%]$  & $\in[1.2\%, 3.2\%]$  &  N/A \\
        \hline
    \end{tabular}
    }
    \label{tab:diff_noise_diff_inj}
\end{table}

This set of tests represents a more combined evaluation, assessing the impact of both noise and injection variations at the same time. The 10 datasets used in this analysis were generated with different injection files and distinct background noise files, as previously outlined in Sec. \ref{sec:test_datasets}. While each individual dataset covers only a portion of the full O3a noise data, 30 days to be exact, the full noise file is used in all datasets collectively. As a result, similarly to Sec. \ref{sec:diff_noise_same_inj}, each dataset still contains some of the approximately 40 real astrophysical signals included in the noise file, with the exact number varying depending on the specific offset used.

Table~\ref{tab:diff_noise_diff_inj} is similar to Tables~\ref{tab:AresGW_same_noise_diff_inj} and~\ref{tab:AresGW_diff_noise_same_inj}. Furthermore, Fig.~\ref{fig:diff_noise_diff_inj} once again presents histograms of the number of injections detected at different values of $\mathcal{M}$ at the three FAR thresholds and exhibits a variance comparable to that shown in Fig.~\ref{fig:diff_noise_same_inj}.

At FAR = 1/month, $\mu_{\rm N}$ was $2797 \pm 207$. The 95\% CI for $\sigma_{\rm N}$ spanned the wide range $[199, 528]$. At FAR = 10/month, $\mu_{\rm N}$ rose to $3909 \pm 129$, with $\sigma_{\rm N} \in [124, 330]$. Finally, at FAR = 100/month, $\mu_{\rm N}$ was $5244 \pm 67$, with a relatively tighter CI for $\sigma_{\rm N}$ of $[64, 170]$. Furthermore, the CIs for $\sigma_{\rm N}/\bar\mu_{\rm N}$ at the three FAR thresholds 1, 10, and 100/month, were $[7.1\%, 18.9\%]$, $[3.2\%, 8.4\%]$, and $[1.2\%, 3.2\%]$, respectively.

\begin{figure}[t]
  \centering
  \includegraphics[width=0.92\linewidth]{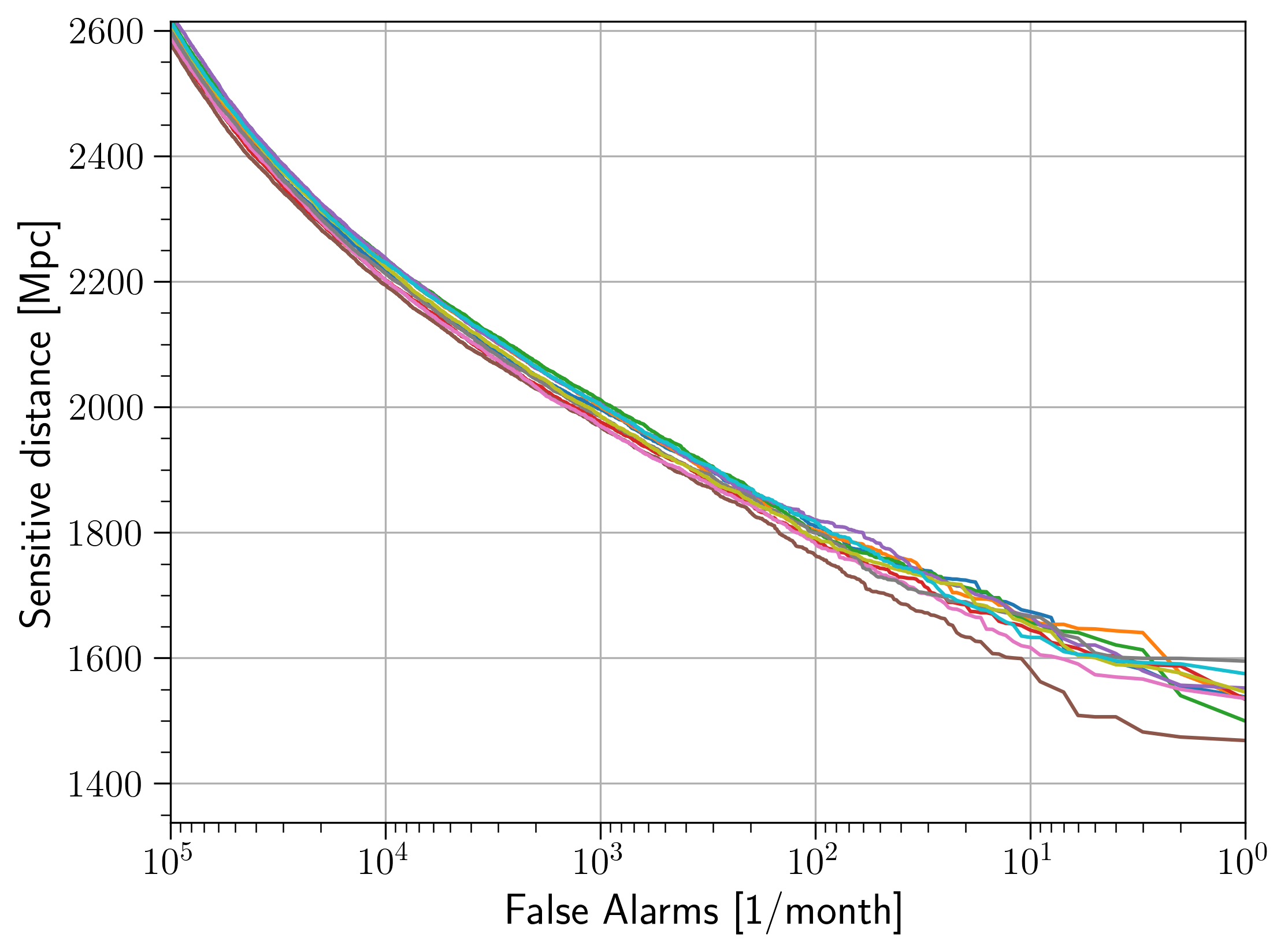}
  \caption{As in Fig. \ref{fig:sensitivity_same_noise_diff_inj} but with various instances of both real detector noise and simulated signal injections. The color scheme is as in Fig. \ref{fig:diff_noise_diff_inj}.} 
  \label{fig:sensitivity_plot_prelimenary_analysis}
\end{figure}

\begin{table}[t]
    \centering
    \renewcommand{\arraystretch}{1.1} 
    \caption{As in Table \ref{tab:sensitivity_distance_same_noise_diff_inj}, but using various instances of both real detector noise and simulated signal injections.}
    \footnotesize
    \resizebox{0.43\textwidth}{!}{
    \begin{tabular}{|r|l|c|c|c|c|}
        \hline
        \rowcolor{blue!15} \# & Offset & Injection/ & $S\text{ (Mpc)}$ & $S\text{ (Mpc)}$  & $S\text{ (Mpc)}$ \\
        \rowcolor{blue!15} \# & (days) & Noise Seed & 1/month & 10/month  & 100/month \\
        \hline
        \rowcolor{blue!10}
        1 & 0 & 2514409456 & 1574.94 & 1632.87 & 1816.48\\
        2 & 0 & 4 & 1468.54 & 1582.38 & 1763.79\\
        \rowcolor{blue!10}
        3 & 5 & 2371 & 1595.04 & 1666.14 & 1799.57\\
        4 & 15 & 12 & 1534.40 & 1644.49 & 1787.31\\
        \rowcolor{blue!10}
        5 & 25 & 7419 & 1545.74 & 1652.03 & 1791.63\\
        6 & 27 & 1213 & 1499.53 & 1655.95 & 1801.17\\
        \rowcolor{blue!10}
        7 & 30 & 102 & 1535.44 & 1662.64 & 1803.80\\
        8 & 35 & 1823 & 1552.18 & 1667.28 & 1820.09\\
        \rowcolor{blue!10}
        9 & 41 & 601 & 1535.54 & 1616.86 & 1781.98\\
        10 & 45 & 1000000 & 1537.94 & 1673.73 & 1809.96\\
        \hline
        \hline
        \rowcolor{blue!10}
        \multicolumn{3}{|c|}{$\mu_{\rm S}$ (mean)} & $1537.9\pm25.2$ & 1645.4$\pm$20.1 & $1797.6\pm12.2$ \\
        \hline
        \rowcolor{blue!10}
        \multicolumn{3}{|c|}{$\sigma_{\rm S}$ (std. dev.)}  &  $ \in [24.3, 64.4]$  & $ \in[19.3, 51.3]$  &  $ \in[11.7, 31.0]$  \\
        \hline
        \rowcolor{blue!10}
        \multicolumn{3}{|c|}{$\sigma_{\rm S}/\bar\mu_{\rm s}$ }  & $\in[1.6\%, 4.2\%]$  & $\in[1.2\%, 3.1\%]$  & $\in[0.7\%, 1.7\%]$  \\
        \hline
    \end{tabular}
    }
    
    \label{tab:sensitivity_distance_diff_noise_diff_inj}
\end{table}

The broad CIs observed, particularly at lower FAR thresholds, clearly reflect \textit{variability in the number of detected injections}. This variability is likely driven primarily by noise fluctuations, as suggested by the results of the previous two subsections. 

Continuing, similarly to the previous subsections, the sensitive distance results at the same three FAR thresholds are presented in Table \ref{tab:sensitivity_distance_diff_noise_diff_inj}, while the sensitive distance as a function of FAR for all tests is shown in Fig. \ref{fig:sensitivity_plot_prelimenary_analysis}.

The CI for $\mu_{\rm S}$ at FAR = 1/month was $1537.9 \pm 25.2$ Mpc, while the CI for $\sigma_{\rm S}$ was $[24.3, 64.4]$ Mpc. At FAR = 10/month, $\mu_{\rm S}$ increased to $1645.4 \pm 20.1$ Mpc. The CI for $\sigma_{\rm S}$ was slightly tighter at $[19.3, 51.3]$ Mpc. Finally, at FAR = 100/month, $\mu_{\rm S}$ reached $1797.6 \pm 12.2$ Mpc, with $\sigma_{\rm S} \in [11.7, 31.0]$ Mpc. Ultimately, the CIs for $\sigma_{\rm S}/\bar\mu_{\rm S}$ at FAR = 1, 10, and 100/month were $[1.6\%, 4.2\%]$ Mpc, $[1.2\%, 3.1\%]$ Mpc, and $[0.7\%, 1.7\%]$ Mpc, respectively.

As observed, the sensitive distance once again proves to be a considerably more robust metric compared to the number of detected injections — particularly when relying on a single test dataset, although such reliance is not ideal. Notably, even at FAR = 1/month, the upper bound of the CI for $\sigma_{\rm S}/\bar\mu_{\rm S}$ is only about 4.2\%, while at $\mathrm{FAR} = 100/\text{month}$, it decreases to approximately 1.7\%. This indicates that, when evaluation is limited to a single test instance, the sensitive distance emerges as the preferred metric. However, it is important to acknowledge that this measure may disadvantage algorithms that do not perform well in detecting high-mass signals, as the sensitive distance is directly influenced by $\mathcal{M}$. This means that, even though the sensitive distance is a very robust metric, it still has its own biases and limitations.

\section{\label{sec:performance_clean} Performance Evaluation on Datasets without Event Contamination}

\begin{figure*}[t]
  \centering
  \includegraphics[width=0.31\textwidth]{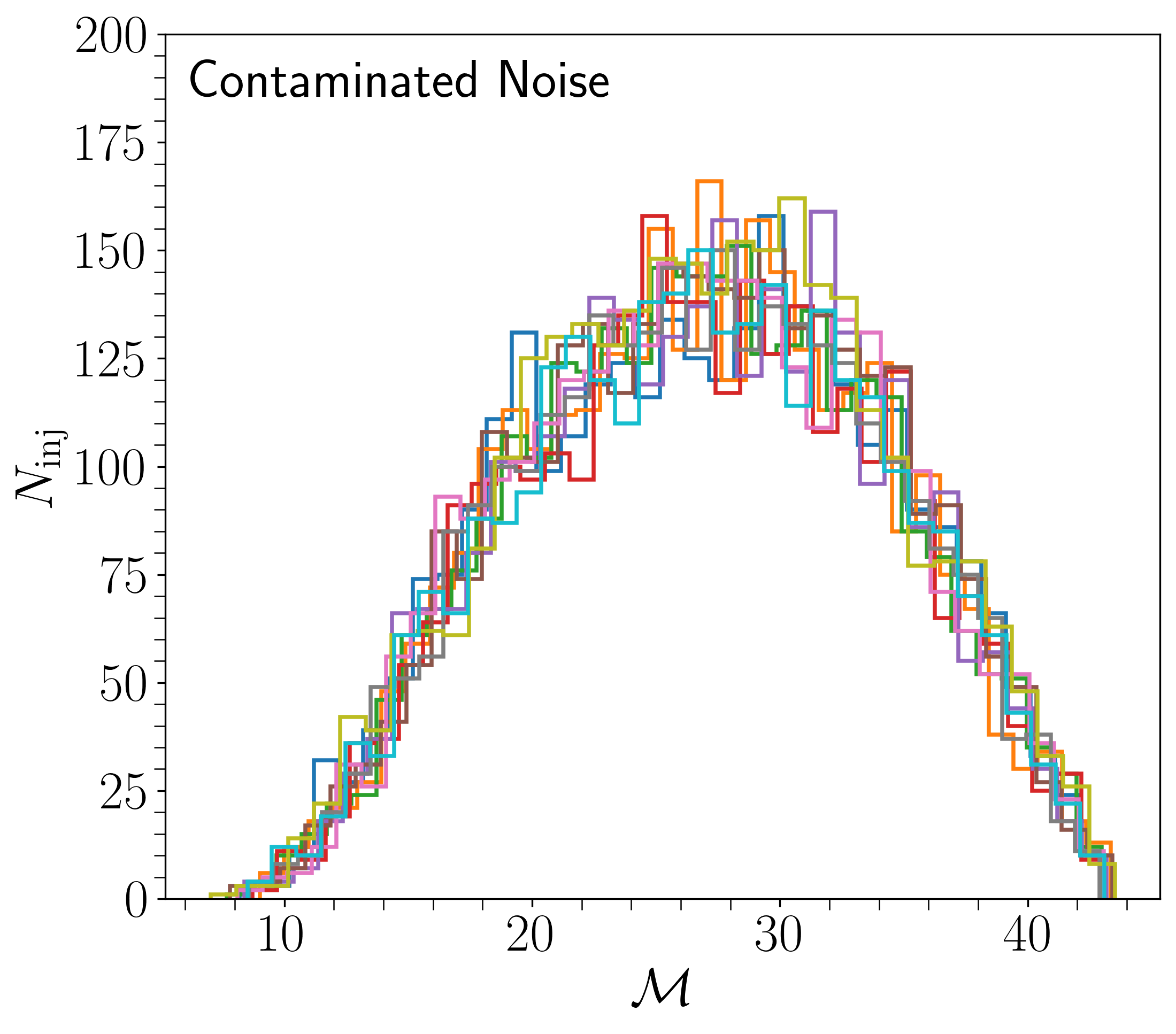} 
   \hspace{0.1cm} 
   \includegraphics[width=0.273\textwidth]{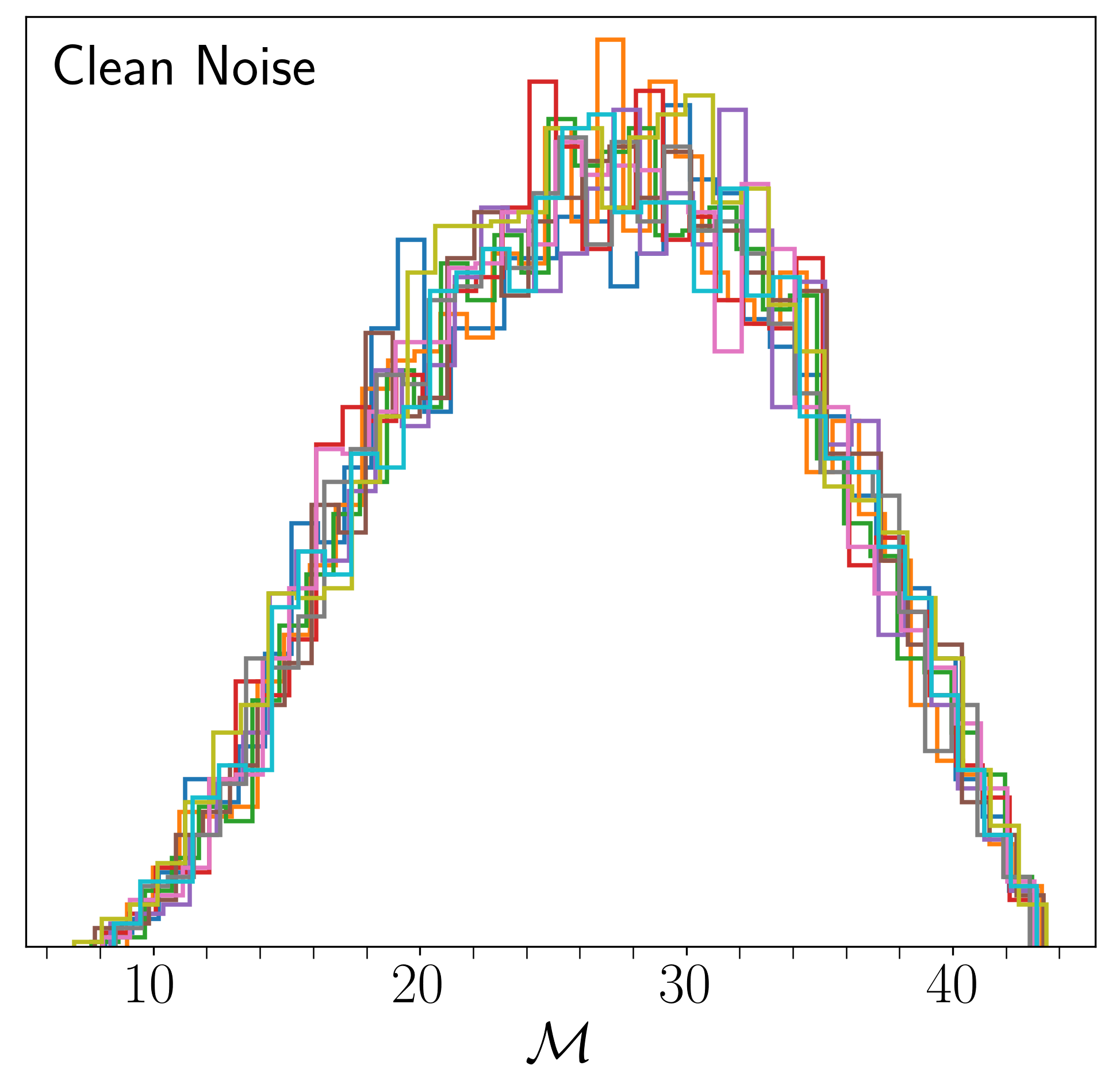}
  \caption{Histogram of various tests at a FAR of 1/month, using identical background noise (offset = 0 and noise seed = 2514409456) cleaned of real GW signals and injected with different sets of signals. The color scheme is as in Fig. \ref{fig:same_noise_diff_inj}.}  \label{fig:same_noise_diff_inj_events_removed}
\end{figure*}

Our objective in this section is to assess the impact of noise contamination that arises from the presence of known GW signals within the noise. To achieve this, we replicate the same analyses described in Sec. \ref{sec:performance}, but this time after removing all known GW signals from the data. Thus, in addition to the GWTC-2 signals, we also exclude all events listed in the AresGW, OGC, GWTC-2.1, and IAS catalogs. The 37 signals that were retained in previous analyses but have now been excluded from the datasets are listed in Table~\ref{tab:appendix_events} in Appendix \ref{appendix_events}. 

Note that many of these events - particularly those originating from the IAS catalog - lie outside the effective training range of our model and are therefore not expected to affect AresGW model 1's results. In contrast, all events included in the AresGW catalog fall within this range. Nevertheless, to ensure a fair comparison between different pipelines and to obtain a more accurate estimate of an algorithm’s performance, all previously known detections should be removed from any test dataset, as we have done here, to avoid introducing additional biases into the evaluation process.

Additionally, since the most important conclusions can be drawn from the category "Datasets with identical noise and varying injections," the results following total event removal for the other two categories — "Datasets with varying noise and identical injections" and "Datasets with both varying noise and injections" — are presented in Appendix~\ref{appendix:clean} for brevity.

Moving forward, we retain only the results for $\mathrm{FAR} = 1/\text{month}$ for both the number of detected injections and the sensitive distance, together with the corresponding tables, as the differences observed at the higher FAR values, 10 and 100/month, were not statistically significant. Consequently, we omit the ${N^{\rm F}}$ histograms for FAR of 10 and 100/month, since their morphology closely mirrors those presented in the previous section. Furthermore, we exclude the sensitive distance plots, as we will later demonstrate that signal contamination has minimal influence on the sensitive distance results, even at FAR = 1/month.

With that being said, we now proceed with the analysis of the \textit{datasets with identical noise and varying injections} (offset = 0 and noise seed = 2514409456), after removing all GW events from the noise. It is important to note that in the corresponding datasets \textit{prior} to event removal, nearly 10 known real GW events were still present in the data, as discussed in Sec.~\ref{sec:same_noise_diff_inj}.

Thus, let us start with the number of detected injections. Fig.~\ref{fig:same_noise_diff_inj_events_removed} shows $N_{1/\text{month}}^F$ as a function of $\mathcal{M}$, both before (left panel) and after (right panel) removal of the GW signals from the noise. Importantly, the difference in $N_{1/\text{month}}^F$ between the two cases is clearly evident. In fact, after removing residual real events from the background data, the number of detected injections in each of the test datasets at a FAR of 1/month increased by approximately \textit{560 injections}.

\begin{table}[t]
    \centering
    \renewcommand{\arraystretch}{1.1} 
    \caption{Number of detected injections and sensitive distance for the AresGW model 1 at FAR = 1/month, evaluated across multiple one-month datasets containing identical real detector noise (offset = 0, noise seed = 2514409456) that has been cleaned of all known GW events.}
    \footnotesize
    \resizebox{0.30\textwidth}{!}{
    \begin{tabular}{|r|c|c|c|}
        \hline
        \rowcolor{blue!15} \# & Injection  & $N^{\rm F}$ & $S \text{ (Mpc)}$  \\
        \rowcolor{blue!15} \# & Seed & 1/month & 1/month   \\
        \hline
        \rowcolor{blue!10}
        1 & 2514409456 & 3475 & 1590.60 \\
        2 & 12019 & 3504 &1600.20 \\
        \rowcolor{blue!10}
        3 & 10209 & 3518 &1572.03 \\
        4 & 9801 & 3494 &1587.90 \\
        \rowcolor{blue!10}
        5 & 6291 & 3433 &1586.44 \\
        6 & 555 & 3458 &1600.90 \\
        \rowcolor{blue!10}
        7 & 291 & 3474 &1598.34 \\
        8 & 93 & 3395 &1587.06 \\
        \rowcolor{blue!10}
        9 & 32 & 3506 &1587.90 \\
        10 & 9 & 3454 &1587.50 \\
        \hline
        \hline
        \rowcolor{blue!10}
        \multicolumn{2}{|c|}{$\mu$ (mean)} & $3471\pm27$ & $1589.9\pm6.1$ \\
        \hline
        \rowcolor{blue!10}
        \multicolumn{2}{|c|}{$\sigma$ (std. dev.)}  & $\in[26, 69]$  & $\in[5.9, 15.5]$ \\
        \hline
        \rowcolor{blue!10}
        \multicolumn{2}{|c|}{$\sigma_{\rm N}/\bar\mu_{\rm N}$ }  & $\in[0.7\%,2.0\%]$  & $\in[0.4\%,1.0\%]$ \\
        \hline
    \end{tabular}
    }
    \label{tab:same_noise_diff_inj_clean}
\end{table}

In particular, at FAR = 1/month, $\mu_{\rm N}$ increased from $2913 \pm 23$ to $3471 \pm 27$, while the 95\% CI for $\sigma_{\rm N}$ remained relatively stable at $[25.9, 68.7]$. The corresponding CI for $\sigma_{\rm N}/\bar{\mu}_{\rm N}$ became $[0.7\%,2.0\%]$.

Notably, the significant boost in detection performance observed for AresGW model 1 appears to be largely driven by the inclusion of the event GW190511\_125545 in the original evaluation dataset. As noted previously, this event was initially identified by its successor, AresGW model 2, but is also detected with a high ranking statistic by model 1. Thus, its presence skewed the FAR estimate, leading to the rejection of many injections at a FAR threshold of 1/month. 

It is also important to remind that the common background file used for all 10 datasets in the "Datasets with identical noise and varying injections" category, prior to event removal, includes GW190511\_125545. This file was not only used in~\cite{challenge1}, but is also the same as used in~\cite{nagarajan2025identifyingmitigatingmachinelearning}. Consequently, the results reported in that study were influenced by the presence of real astrophysical events, affecting the validity of the performance evaluation. For example,~\cite{nagarajan2025identifyingmitigatingmachinelearning} reported that AresGW model 1 detected 2892 injections at a FAR of 1/month. However, when we re-evaluated the dataset with real events removed, the same model detects 3475 injections\footnote{These results refer to the first line of Table~\ref{tab:same_noise_diff_inj_clean}.} -highlighting a substantial difference. Therefore, to ensure fair and reliable comparisons between algorithms, all known events must be excluded from evaluation datasets. Our complete recommendations for robust benchmarking practices are detailed in Sect.~\ref{sec:fair_comparisons}.

However, the sensitive distance tells a different story. Following removal of real events, $\mu_S$ increased only slightly — from $1574.6 \pm 6.5$ Mpc to $1589.9\pm6.1$  Mpc. The new CIs for $\sigma_{\rm S}$ and $\sigma_{\rm S}/\bar{\mu}_S$ are then $[5.9, 15.5]$ Mpc and $[0.4\%,1.0\%]$ Mpc, respectively. 

These results indicate that the sensitive distance remains a robust performance metric even when the background noise contains real signals. In fact, the sensitive distance values remained consistent even after the inclusion of approximately 600 additional detected injections.

The values of $N_{1/\text{month}}^F$ and $S_{1/\text{month}}$ for all datasets in this category are presented in Table~\ref{tab:same_noise_diff_inj_clean}.

\section{\label{sec:fair_comparisons} Towards Reliable and Robust Evaluation of Gravitational-Wave Detection Algorithms}

In the preceding analyses, we highlighted several key issues that currently affect the evaluation of GW detection algorithms. These include the inadvertent inclusion of real GW events within noise datasets and the unreliability of performance metrics — such as the number of detected injections at a fixed FAR — when derived from limited amounts of data. Such issues represent only a subset of the factors that can introduce bias or lead to misleading conclusions about the effectiveness of a detection pipeline. This raises a critical question: How can we address and mitigate these challenges to ensure more reliable and accurate evaluation frameworks?

As we discussed in depth before, one key consideration is the variability introduced by different noise realizations and injection configurations. Evaluating algorithms on a single or limited set of datasets can produce results that are highly sensitive to specific conditions, rather than reflecting generalizable performance. To mitigate this, we recommend evaluating each model across a large number of independently sampled datasets, with varying noise and injection seeds and offsets. This approach enables the calculation of reliable summary statistics — such as the mean, variance, coefficient of variation, and their respected CIs — for each metric of interest. Without these, comparisons may appear definitive, while being influenced by random fluctuations or the inclusion of particularly favorable (or unfavorable) test samples. 

Another important aspect is the integrity of the background data. The presence of real GW signals in the test set can introduce unintended biases, since different algorithms are more or less sensitive to specific events. This can result in decreased detection rates that do not accurately reflect performance on clean noise, as we demonstrated for the AresGW model 1 in Sec. \ref{sec:performance_clean}. Therefore, to enable a controlled and unbiased comparison of pipelines or a reliable evaluation of one code, it is critical to remove all known real events from the background data prior to evaluation, as we did in Sec. \ref{sec:performance_clean}.

In addition, the use of multiple metrics is essential because each metric carries its own inherent biases. For example, the sensitive distance may favor algorithms that perform better at higher $\mathcal{M}$ due to its dependence on it. Thus, incorporating multiple evaluation metrics provides a more comprehensive understanding of where and when an algorithm excels, and also enables fairer comparisons across different algorithms.

Finally, the use of standardized datasets and evaluation protocols can further improve the reproducibility and transparency of algorithm comparisons. For this reason, we provide here the noise and injection seeds and the noise offsets. Furthermore the lists of real event times in the different datasets and the injection files used for the set of datasets analyzed in Sec.~\ref{sec:diff_noise_same_inj} are publicly available at \texttt{\url{https://gitlab.com/Alexandra1120/aresgw-variance}}. We recommend that algorithm evaluations be performed following the same procedure described in Section~\ref{sec:performance_clean}, using the same datasets with all known GW signals removed. Thus, by establishing common benchmarks and protocols, the community can better assess progress and identify real advances in detection methodologies.

\section{\label{sec:results} Conclusions}

In this work, we conducted a comprehensive statistical evaluation of AresGW model 1 — a machine learning-based GW detection algorithm — under conditions representative of real-world deployment. By analyzing its sensitivity across 28 different one-month datasets, we quantified the robustness of its performance using two metrics: the number of detected injections and the sensitive distance, at various FAR thresholds.

Our findings reveal several key insights. First, detection variability is significantly influenced by noise fluctuations rather than injection configurations. Even when the injection parameters remain constant, changes in detector noise produce wide variances in detection counts — particularly at lower FARs such as 1/month. In contrast, variations in the injection set (with fixed noise) result in far more stable performance metrics, suggesting that the model generalizes well across the injection parameter space.

Second, we demonstrate that the sensitive distance is a more stable and reliable performance metric than the raw number of detected injections. Even at low FAR thresholds, where performance is most variable, the sensitive distance exhibited notably narrower CIs. This makes it preferable for benchmarking and comparisons, especially when only a small number of test datasets are available.

Third, we identified the significant impact of dataset contamination by real GW events. In particular, a single high-ranking event — GW190511\_125545 — was shown to substantially bias FAR estimates and lead to under-reporting of injection recovery rates. Once this and other known signals were removed, the model’s performance, when measured in terms of detected injections at a fixed FAR, improved markedly. This impacts previous results drawn from contaminated datasets (e.g. \cite{nagarajan2025identifyingmitigatingmachinelearning}), and emphasizes the need for clean datasets in reliable model comparisons.

Finally, our work highlights the pitfalls of relying on single-dataset evaluations. We found that performance metrics derived from a single one-month dataset can be highly unrepresentative at FAR of 1/month, due to statistical variance. Thus, to ensure a rigorous and reliable evaluation, we recommend the use of multiple independent datasets, reporting of statistical uncertainty (mean, variance, coefficient of variation, CIs), and the use of both injection count and sensitive distance as complementary metrics.

\section*{ACKNOWLEDGMENTS}

We are grateful to Melissa Lopez and Narenraju Nagarajan for useful discussions. Additionally, we acknowledge the support
provided by the IT Center of the Aristotle University of
Thessaloniki (AUTh) as our results have been produced
using the AUTh High-Performance Computing Infrastructure and Resources. We also acknowledge support
from COST Action (European Cooperation in Science
and Technology) CA17137 (G2Net) and the EU Horizon project nr. 101131928 (ACME). 
This research has made use of data or software obtained from the Gravitational Wave Open Science Center (gw-openscience.org), a service of LIGO Laboratory,
the LIGO Scientific Collaboration, the Virgo Collaboration, and KAGRA. LIGO Laboratory and Advanced
LIGO are funded by the United States National Science Foundation (NSF) as well as the Science and Technology Facilities Council (STFC) of the United Kingdom, the Max-Planck-Society (MPS), and the State of
Niedersachsen/Germany for support of the construction
of Advanced LIGO and construction and operation of
the GEO600 detector. Additional support for Advanced
LIGO was provided by the Australian Research Council. Virgo is funded, through the European Gravitational Observatory (EGO), by the French Centre National de Recherche Scientifique (CNRS), the Italian Istituto Nazionale di Fisica Nucleare (INFN) and the Dutch
Nikhef, with contributions by institutions from Belgium,
Germany, Greece, Hungary, Ireland, Japan, Monaco,
Poland, Portugal, Spain. The construction and operation of KAGRA are funded by Ministry of Education,
Culture, Sports, Science and Technology (MEXT), and
Japan Society for the Promotion of Science (JSPS), National Research Foundation (NRF and Ministry of Science and ICT (MSIT) in Korea, Academia Sinica (AS)
and the Ministry of Science and Technology (MoST) in
Taiwan.

\appendix

\section{Performance Evaluation on Datasets Without Event Contamination: Remaining Categories}
\label{appendix:clean}

Here, we repeat the analyses presented in Secs.~\ref{sec:diff_noise_same_inj} and~\ref{sec:diff_noise_diff_inj}, this time after removing the remaining events listed in Table~\ref{tab:appendix_events}. This serves as a complementary analysis to that of Sec.~\ref{sec:performance_clean}.

\subsection{Varying Clean Noise and Identical Injections}
\label{sec:diff_noise_same_inj_clean}

Considering the case of different background files with identical injections (in terms of source parameters), the increase in $\mu_{\rm N}$ appears much smaller compared to Sec.~\ref{sec:performance_clean}. Specifically, $\bar{\mu}_{\rm N}$ increased statistically insignificantly from 2634 to 2693. Thus, after event removal, the 95\% CI for $\mu_{\rm N}$ is given by $2693 \pm 445$. However, the variance in performance appears to have increased, with the 95\% CIs for $\sigma_{\rm N}$ and $\sigma_{\rm N} / \bar{\mu}_{\rm N}$ being [427, 1134] and [19.0\%, 50.4\%], respectively.

\begin{figure}[t]
  \centering
  \includegraphics[width=0.93\linewidth]{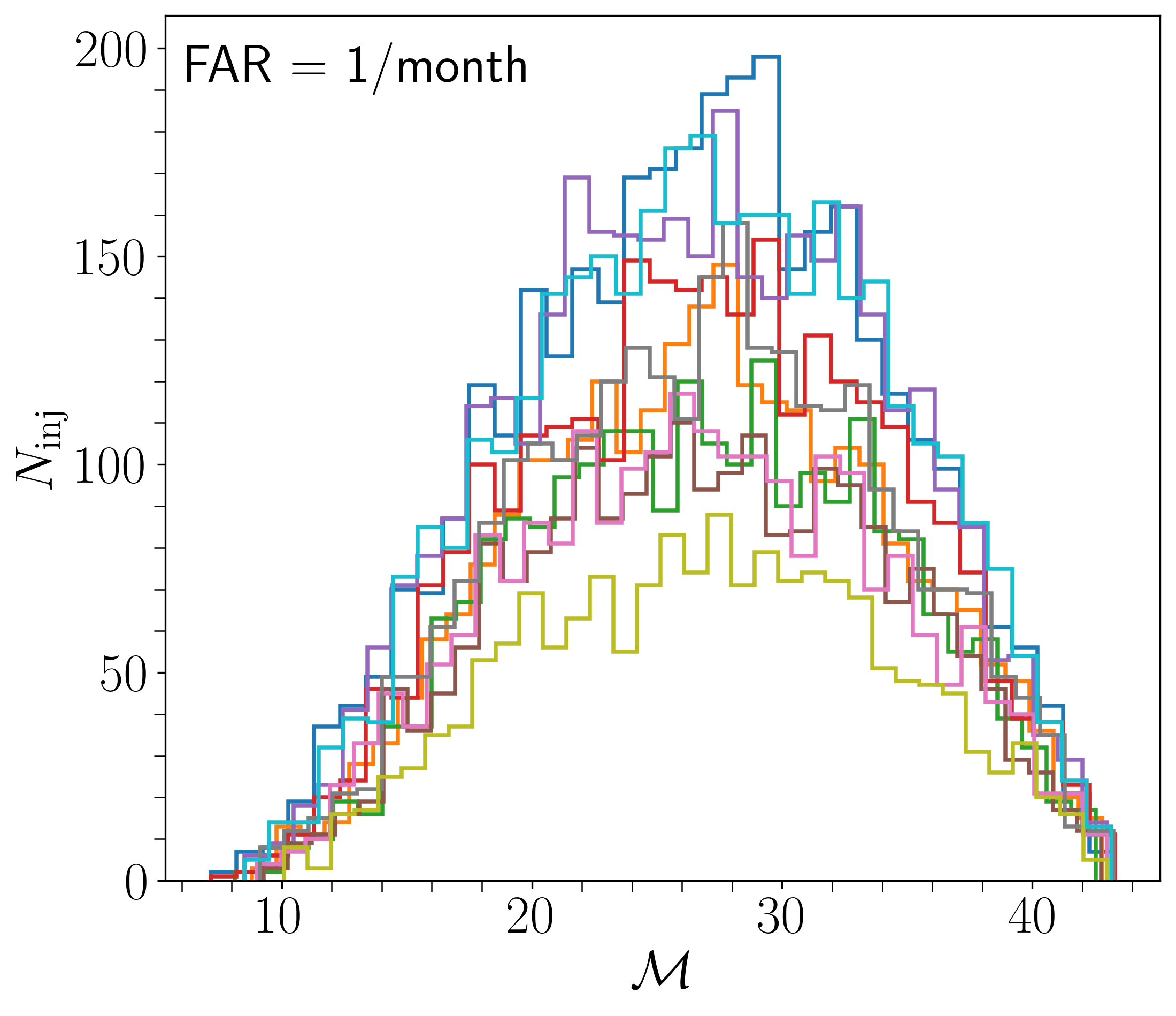}
  \caption{As in Fig. \ref{fig:same_noise_diff_inj_events_removed}, but using different real clean detector noise and a fixed set of simulated signal injections. The color scheme is as in Fig. \ref{fig:diff_noise_same_inj}.}  \label{fig:diff_noise_same_inj_events_removed}
\end{figure}

\begin{table}[t]
    \centering
    \renewcommand{\arraystretch}{1.1} 
    \caption{As in Table \ref{tab:same_noise_diff_inj_clean}, but using different real clean detector noise and a fixed set of simulated signal injections.}
    \renewcommand{\arraystretch}{1.1} 
    \footnotesize
    \resizebox{0.41\textwidth}{!}{
    \begin{tabular}{|r|l|c|c|c|}
        \hline
        \rowcolor{blue!15} \# & Offset (days) & Noise Seed & $N_{\rm{1/month}}^F$ & $S_{1{\rm/month}}$ (Mpc)\\
        \hline
        \rowcolor{blue!10}
        1 & 0 & 2514409456 & 3475 &1590.60 \\
        2 & 0 & 500724 & 2597 &1505.19 \\
        \rowcolor{blue!10}
        3 & 0 & 16 & 2733 &1472.77 \\
        4 & 10 & 145 & 1668 &1458.94 \\
        \rowcolor{blue!10}
        5 & 20 & 3199 & 2307 &1515.50 \\
        6 & 20 & 313 & 2191 &1475.02 \\
        \rowcolor{blue!10}
        7 & 30 & 1009 & 2788 &1464.98 \\
        8 & 30 & 2 & 2242 &1554.19 \\
        \rowcolor{blue!10}
        9 & 40 & 6 & 3470 &1572.54 \\
        10 & $\approx$46.3 & 7897 & 3456 & 1568.46\\
        \hline
        \hline
        \rowcolor{blue!10}
        \multicolumn{3}{|c|}{$\mu$ (mean)}  & $2693\pm445$  & $1517.8\pm35.7$ \\
        \hline
        \rowcolor{blue!10}
        \multicolumn{3}{|c|}{$\sigma$ (std. dev.)}  & $\in[427, 1134]$  & $\in[34.4, 91.2]$ \\
        \hline
        \rowcolor{blue!10}
        \multicolumn{3}{|c|}{$\sigma/\bar\mu$ }  & $\in[19.0\%,50.4\%]$  & $\in[2.3\%, 6.0\%]$ \\
        \hline
    \end{tabular}
    }
    \label{tab:diff_noise_same_inj_clean}
\end{table}

But why is this the case? Why do we observe only a small difference in $\bar{\mu}_{\rm N}$, yet a substantial increase in the variability of performance? A total of 37 real GW events were collectively removed from all datasets. However, as mentioned previously, only the event GW190511\_125545 appears to significantly affect the performance of the AresGW model~1 at a FAR of 1/month. This implies that datasets not containing this event showed little to no change in performance after event removal, while those that did include it \textit{can} exhibit a marked increase in detected injections.

However, even the effect of GW190511\_125545 is not guaranteed: due to timeshifts applied to the Livingston data, it is possible that the event is not combined with noise exhibiting similar characteristics to the original instance, in which case the network does not assign a high ranking statistic. As a result, although the average number of detections remained nearly the same, the increased spread in performance across datasets led to higher observed variance. In contrast, in Sec.~\ref{sec:performance_clean}, all datasets shared the same background, which originally included GW190511\_125545, paired with a noise segment in the second detector that resulted in a high ranking statistic. Consequently, performance was compromised across all of these datasets.

\begin{figure}[t]
  \centering
  \includegraphics[width=0.93\linewidth]{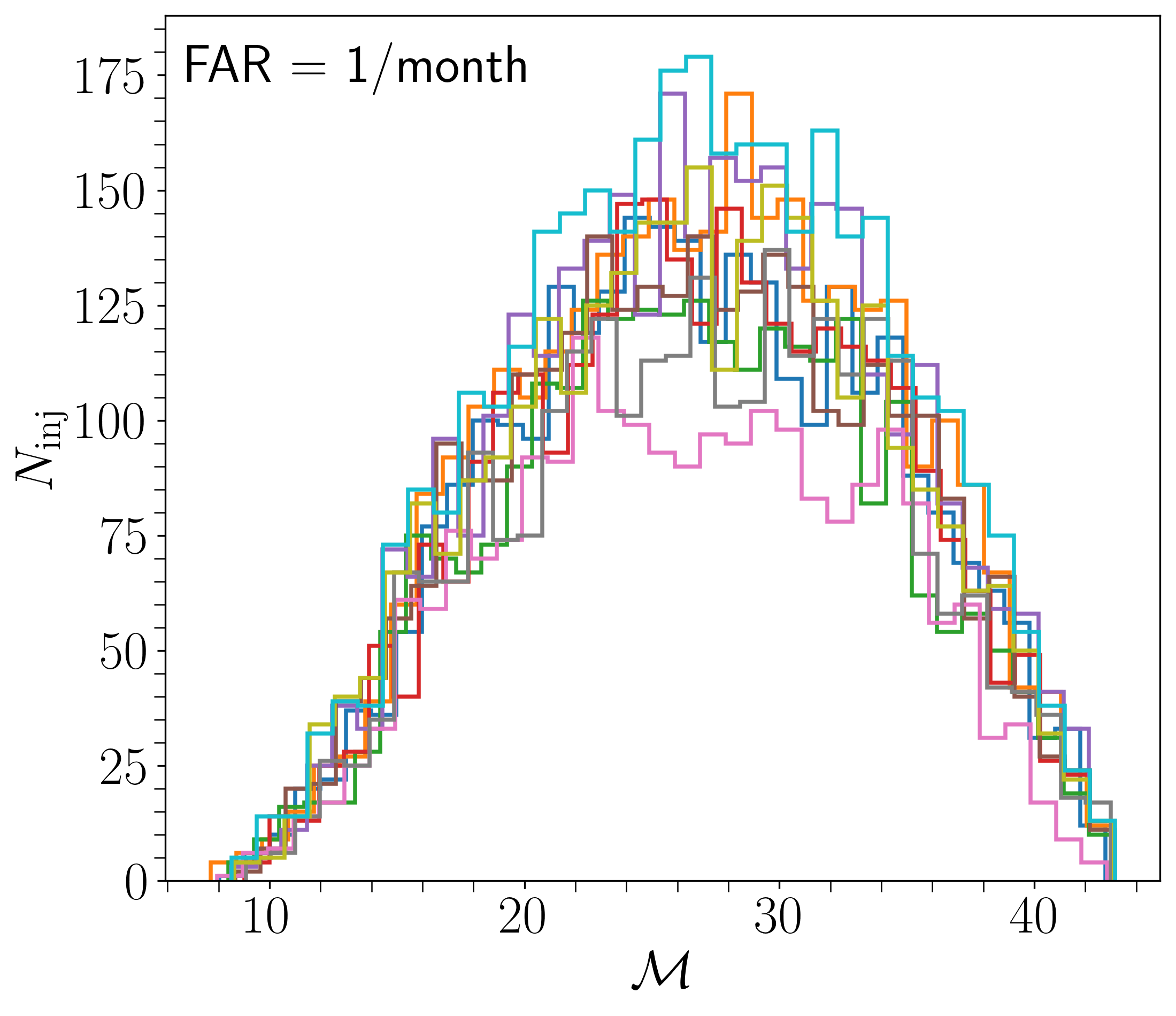}
  \caption{As in Fig. \ref{fig:same_noise_diff_inj_events_removed}, but using various instances of both real detector noise and simulated signal injections. The color scheme is as in Fig. \ref{fig:diff_noise_diff_inj}.}  \label{fig:diff_noise_diff_inj_events_removed}
\end{figure}

\begin{table}[t]
    \centering
    \renewcommand{\arraystretch}{1.1} 
    \caption{As in Table \ref{tab:same_noise_diff_inj_clean}, but using various instances of both real detector noise and simulated signal injections.}
    \footnotesize
    \resizebox{0.455\textwidth}{!}{
    \begin{tabular}{|r|l|c|c|c|}
        \hline
        \rowcolor{blue!15} \# & Offset (days) & Noise/Injection Seed & $N_{\rm{1/month}}^F$ & $S_{1{\rm/month}} (Mpc)$ \\
        \hline
        \rowcolor{blue!10}
        1 & 0 & 2514409456 &3475  &1590.60  \\
        2 & 0 & 4 &2863  &1599.10 \\
        \rowcolor{blue!10}
        3 & 5 & 2371 &2618  &1595.04 \\
        4 & 15 & 12 &2847  &1534.12 \\
        \rowcolor{blue!10}
        5 & 25 & 7419 &2970  &1545.74 \\
        6 & 27 & 1213 &2566  &1499.53 \\
        \rowcolor{blue!10}
        7 & 30 & 102 &3051  &1535.44  \\
        8 & 35 & 1823 &3181  &1552.18  \\
        \rowcolor{blue!10}
        9 & 41 & 601 &2158  & 1535.54\\
        10 & 45 & 1000000 &2821  & 1537.94 \\
        \hline
        \hline
        \rowcolor{blue!10}
        \multicolumn{3}{|c|}{$\mu$ (mean)}  & $2855\pm258$  & $1552.5\pm23.1$ \\
        \hline
        \rowcolor{blue!10}
        \multicolumn{3}{|c|}{$\sigma$ (std. dev.)}  & $\in[248, 658]$  & $\in[22.2, 59.0]$ \\
        \hline
        \rowcolor{blue!10}
        \multicolumn{3}{|c|}{$\sigma_/\bar\mu$ }  & $\in[9.7\%,23.0\%]$  & $\in[1.4\%,3.8\%]$ \\
        \hline
    \end{tabular}
    }
    \label{tab:diff_noise_diff_inj_clean}
\end{table}

Turning to the sensitive distance, the increase in $\bar{\mu}_S$ remains small while the variance continues to have tight CIs—once again demonstrating its robustness — rising from 1512.7~Mpc to 1517.8~Mpc following the removal of real events. Thus, the updated CI for $\mu_{\rm S}$ is $1517.8\pm35.7$ Mpc, while the CIs for $\sigma_{\rm S}$ and $\sigma_{\rm S}/\bar\mu_{\rm S}$ remain relatively narrow, spanning [34.4, 91.2] and [2.3\%, 6.0\%], respectively.

Fig.~\ref{fig:diff_noise_same_inj_events_removed} is analogous to Fig.~\ref{fig:same_noise_diff_inj_events_removed}, but corresponds to this dataset category. Similarly, Table~\ref{tab:diff_noise_same_inj_clean} adopts the same structure as Table~\ref{tab:same_noise_diff_inj_clean}.

\subsection{\label{sec:diff_noise_diff_inj_clean} Both Varying Clean Noise and Injections}

Finally, we examine the impact of removing known GW signals from datasets varying both in background noise and injections. Fig. \ref{fig:diff_noise_diff_inj_events_removed} 
and Table \ref{tab:diff_noise_diff_inj_clean} present the same information for this dataset category as Fig. \ref{fig:same_noise_diff_inj_events_removed} and Table \ref{tab:same_noise_diff_inj_clean}, respectively.

Here, in terms of the number of injections detected, at FAR = 1/month, $\bar{\mu}_{\rm N}$  increased from 2797 to 2855, with a 95\% CI for $\mu_{\rm N}$ of $2855\pm258$, with $\sigma_{\rm N}$ $\in[248, 658]$. The corresponding CI for $\sigma_{\rm N}/\bar\mu_{\rm N}$ is $[9.7\%,23.0\%]$.

Note that, as mentioned in Appendix \ref{sec:diff_noise_same_inj_clean}, the difference in $\bar{\mu}_{\rm N}$ is not necessarily indicative of a consistent performance difference, between individual datasets, after the extraction of the remaining GW signals. The difference in $\bar{\mu}_{\rm N}$ before and after event removal appears relatively small because most real events do not significantly impact the model's behavior at $\mathrm{FAR} = 1/\text{month}$. However, as previously mentioned, for datasets containing the event \texttt{GW190511\_125545}, the performance of AresGW model~1 can be notably degraded if the event is not removed, as demonstrated in Sec.~\ref{sec:performance_clean}.

We now repeat the same analysis using the sensitive distance and, as before, find that, unlike the number of detected injections, the sensitive distance is affected across all datasets only to a statistically insignificant degree. At $\mathrm{FAR} = 1/\text{month}$, $\bar{\mu}_{\rm S}$ is 1552.5 Mpc, which is close to the previous value of 1537.9 Mpc. Thus, the CI for $\mu_{\rm S}$, after the removal of the remaining GW events, is $1552.5\pm23.1$ Mpc. Furthermore, the CIs for $\sigma_{\rm S}$ and $\sigma_{\rm S}/\bar\mu_{\rm S}$ are $[22.2, 59.0]$ Mpc and $[1.4\%,3.8\%]$, respectively.

\section{Real Signals in Noise}

\begin{table}[b]
    \centering
    \renewcommand{\arraystretch}{1.1} 
    \caption{List of all GW events present in the noise file from \cite{challenge1} and the first catalog they were listed in.}
    \footnotesize
    \resizebox{0.225\textwidth}{!}{
    \begin{tabular}{|r|c|c|c|}
\hline
\rowcolor{blue!15}\# & Name & Catalog \\
\hline
\rowcolor{blue!10}
1 & GW190403\_051519 & GWTC-2.1\\
2 & GW190404\_142514 & OGC-3\\
\rowcolor{blue!10}
3 & GW190426\_082124 & AresGW\\
4 & GW190426\_190642 & GWTC-2.1\\
\rowcolor{blue!10}
5 & GW190427\_180650 & OGC-3\\
6 & GW190511\_125545 & AresGW\\
\rowcolor{blue!10}
7 & GW190511\_163209 & IAS-HM\\
8 & GW190523\_085933 & AresGW\\
\rowcolor{blue!10}
9 & GW190524\_134109 & IAS-HM\\
10 & GW190530\_030659 & IAS-HM\\
\rowcolor{blue!10}
11 & GW190604\_103812 & IAS-HM\\
12 & GW190605\_025957 & IAS-HM\\
\rowcolor{blue!10}
13 & GW190607\_083827 & AresGW\\
14 & GW190614\_134749 & AresGW\\
\rowcolor{blue!10}
15 & GW190615\_030234 & IAS-HM\\
16 & GW190701\_203306 & GWTC-2\\
\rowcolor{blue!10}
17 & GW190704\_104834 & IAS-3\\
18 & GW190705\_164632 & AresGW\\
\rowcolor{blue!10}
19 & GW190706\_222641 & GWTC-2\\
20 & GW190707\_093326 & GWTC-2\\
\rowcolor{blue!10}
21 & GW190718\_160159 & IAS-3\\
22 & GW190725\_174728 & OGC-3\\
\rowcolor{blue!10}
23 & GW190805\_105432 & OGC-3\\
24 & GW190805\_211137 & GWTC-2.1\\
\rowcolor{blue!10}
25 & GW190806\_033721 & IAS-HM\\
26 & GW190814\_192009 & IAS-3\\
\rowcolor{blue!10}
27 & GW190818\_232544 & IAS-3\\
28 & GW190821\_124821 & IAS-3\\
\rowcolor{blue!10}
29 & GW190904\_104631 & AresGW\\
30 & GW190906\_054335 & IAS-3\\
\rowcolor{blue!10}
31 & GW190910\_012619 & IAS-3\\
32 & GW190911\_195101 & IAS-HM\\
\rowcolor{blue!10}
33 & GW190916\_200658 & OGC-3\\
34 & GW190917\_114629 & GWTC-2.1\\
\rowcolor{blue!10}
35 & GW190920\_113516 & IAS-3\\
36 & GW190925\_232845 & OGC-3\\
\rowcolor{blue!10}
37 & GW190926\_050336 & OGC-3\\
\hline
\label{tab:appendix_events}
\end{tabular}}
\end{table}

\label{appendix_events}
Table~\ref{tab:appendix_events} lists all known events in the O3a data that are not part of the GWTC-2 catalog but were still present in the noise used for MLGWSC-1 and subsequent studies that adopted it as a baseline. For each event, we provide the event name and the first catalog in which it was reported.

\section{\label{appendix:statistics} Statistics}

As discussed in Sec. \ref{sec:intro}, our objective is to perform a statistical analysis of AresGW model 1 to demonstrate its capabilities and, crucially, to highlight the limitations of single-month evaluations. To achieve this, we require a robust measure of confidence for each performance metric. For example, instead of simply providing the sample mean number of detected injections ($\bar{\mu}_{\rm N}$) at a specific FAR value, a more effective approach is to compute the 95\% CI for the mean $\mu_{\rm N}$ and repeat this process across different FAR thresholds. Additionally, to further emphasize the inconsistency of results from single-dataset evaluations, we computed the CIs for the standard deviation across different FAR thresholds, providing a more comprehensive assessment of statistical fluctuations. In this section, we outline both the parametric and non-parametric methodologies for computing all the necessary CIs and specify the conditions under which each method is used.

\subsection{Confidence Interval Estimation}
A point estimate $\hat{\theta}$ derived from a sample \textit{provides no information about the accuracy} of the estimation of the parameter $\theta$. The estimate $\hat{\theta}$ we obtain from a sample is a single value, and thus we do not know how close it is to the true value of $\theta$ - in our context $\theta$ corresponds to the true values of $\mu_{\rm N}$, $\sigma^2_{\rm N}$ etc. Moreover, this estimate varies depending on the sample.

For instance, consider the sample mean $\bar{x}$ calculated from a random sample of size $n$. If we were to take a different random sample of the same size, the value of $\bar{x}$ would likely change - it might be closer to, or further from, the true population mean $\mu$ than the previous estimate.

This variability is why, in addition to reporting a point estimate $\hat{\theta}$, it is important to calculate an interval $[\theta_1, \theta_2]$ that contains the true value of the parameter $\theta$ with a certain probability $1 - \alpha$. In other words, there is a probability $\alpha$ that the interval does not contain the true value — this $\alpha$ is referred to as the error probability.

In the following calculations, we assume that the observations \(x_1, \ldots, x_n\) are independent. As we mentioned before, we will introduce both parametric and non-parametric approaches. Note that the parametric approach for computing CIs requires knowledge of the sampling distribution of the estimator (e.g., \(\bar{x}\) or \(s^2\))\footnote{The estimators \(\bar{x}\) and \(s^2\) are the best estimates of \(\mu\) and \(\sigma^2\), respectively, according to the maximum likelihood method.}.

\subsection{Confidence Interval of the Mean \texorpdfstring{$\mu$}{mu} under Normality}

The CI for the population mean $\mu$ is constructed based on the distribution of the sample mean $\bar{x}$. The sample mean $\bar{x}$ serves as an unbiased estimator of $\mu$, as its mean is $\mu$. The variance of $\bar{x}$ is derived as follows:

\begin{equation}
\sigma_{\bar{x}}^2 \equiv \operatorname{Var}[\bar{x}] = \operatorname{Var}\left[\frac{1}{n} \sum_{i=1}^n x_i\right]= \frac{\sigma^2}{n},
\end{equation}
where $\sigma^2$ denotes the variance of the underlying random variable $\theta$. The standard deviation of $\bar{x}$, often referred to as the standard error of the mean, is therefore:

\begin{equation}
\sigma_{\bar{x}} = \frac{\sigma}{\sqrt{n}}.
\end{equation}

The sampling distribution of $\bar{x}$ depends both on the population distribution and the sample size $n$. If the population distribution is normal, i.e., $\theta \sim N(\mu, \sigma^2)$, then $\bar{x}$ is also normally distributed. Even when the population distribution is unknown, the Central Limit Theorem ensures that for sufficiently large $n$ (typically $n \geq 30$), the distribution of $\bar{x}$ approximates normality. Therefore, in both cases:

\begin{equation}
\bar{x} \sim N\left(\mu, \frac{\sigma^2}{n}\right).
\end{equation}

Nevertheless, in practice, the population $\sigma^2$ is generally unknown. A natural approach is to estimate it using the sample variance $s^2$. However, substituting $s^2$ for $\sigma^2$ introduces additional uncertainty, and, as a result, the sampling distribution of $\bar{x}$ is no longer exactly normal. Instead, the standardized statistic:

\begin{equation}
t = \frac{\bar{x} - \mu}{s / \sqrt{n}},
\end{equation}
follows a \textit{Student’s t-distribution} with $n - 1$ degrees of freedom. This distribution is symmetric, but has heavier tails than the normal distribution, accounting for the additional uncertainty introduced by estimating $\sigma^2$. The degrees of freedom, $n-1$, reflect the fact that one degree of freedom is lost in the estimation of $s^2$ from the sample. For large $n$, the t-distribution converges to the standard normal distribution and the difference between them becomes negligible.

To construct a $(1 - \alpha) \times 100\%$ CI for $\mu$, we rely on the following probabilistic statement for the standard normal distribution:
\begin{equation}
\begin{split}
\mathrm{P}\left(-z_{1-\alpha / 2}\leq z \leq z_{1-\alpha / 2}\right) &= \\
\Phi\left(z_{1-\alpha / 2}\right) - \Phi\left(-z_{1-\alpha / 2}\right) &= 1 - \alpha,
\end{split}
\label{eq:possibility}
\end{equation}
where $\Phi$ denotes the cumulative distribution function (CDF) of the standard normal distribution. Replacing $z$ with the $t$-statistic and solving the inequality in Eq. (\ref{eq:possibility}) for $\mu$, the CI becomes:
\begin{equation}
\left[\bar{x} - t_{n-1, 1 - \alpha/2} \cdot \frac{s}{\sqrt{n}}, \quad \bar{x} + t_{n-1, 1 - \alpha/2} \cdot \frac{s}{\sqrt{n}}\right],
\end{equation}
which provides the $(1 - \alpha)$ CI for $\mu$.  In our analysis, we set $\alpha = 0.05$, which yields a CI of 95\%. Furthermore, note that the quantity $t_{n-1, 1 - \alpha/2} \cdot \frac{s}{\sqrt{n}}$ is commonly referred to as the margin of error of the CI for $\bar{x}$.

\subsection{Confidence Interval of the Variance \texorpdfstring{$\sigma^2$}{sigma} under Normality}

To construct a CI for the population $\sigma^2$ using the parametric approach, we assume, again, that the underlying population is normally distributed. Given a sample of size $n$ with observations $x_1, x_2, \dots, x_n$, the sample variance is computed as:
\begin{equation}
s^2 = \frac{1}{n - 1} \sum_{i=1}^{n} (x_i - \bar{x})^2 .
\label{eq:s}
\end{equation}

Thus, under the assumption of normality, the statistic
\begin{equation}
\chi^2 = \frac{(n - 1)s^2}{\sigma^2},
\end{equation}
follows a \textit{chi-square distribution} with $n - 1$ degrees of freedom.  Here, one degree of freedom is lost due to the constraint imposed by the sample mean $\bar{x}$. 

This forms the basis for constructing a CI for $\sigma^2$. Specifically, the probability statement for the chi-square distribution is:
\begin{equation}
\mathrm{P}\left(\chi^2_{1 - \alpha/2} \leq \frac{(n - 1)s^2}{\sigma^2} \leq \chi^2_{\alpha/2} \right) = 1 - \alpha,
\label{eq:probability_x}
\end{equation}
where $\chi^2_{\alpha/2}$ and $\chi^2_{1 - \alpha/2}$ are the critical values of the chi-square distribution corresponding to the lower and upper $\alpha/2$ quantiles, respectively.

Solving the inequality in Eq. (\ref{eq:probability_x}) for $\sigma^2$ gives the $(1 - \alpha) \times 100\%$ CI for the population variance:
\begin{equation}
\left[ \frac{(n - 1)s^2}{\chi^2_{1 - \alpha/2}}, \quad \frac{(n - 1)s^2}{\chi^2_{\alpha/2}} \right].
\end{equation}

If a CI for the population standard deviation $\sigma$ is desired instead, it can be obtained by taking the square root of the bounds:
\begin{equation}
\left[ \sqrt{ \frac{(n - 1)s^2}{\chi^2_{1 - \alpha/2}} }, \quad \sqrt{ \frac{(n - 1)s^2}{\chi^2_{\alpha/2}} } \right].
\end{equation}
As in the mean, we will use a significance level of $\alpha = 0.05$, corresponding to a 95\% CI in our analysis.

\subsection{Non-Parametric Confidence Intervals for Non-Normal Sample Distributions}
\label{sec:bootstrap}

What happens when the sample we want to analyze does not follow a normal distribution? In that case, to estimate confidence intervals (CIs) for the mean ($\mu$), standard deviation ($\sigma$), and variance ($\sigma^2$) without assuming normality, we employed the \textit{non-parametric bootstrap method}. This approach is particularly useful for small sample sizes and unknown or non-normal underlying distributions.

Given our small sample size of $n = 10$, we generated $B = 10{,}000$ bootstrap replicates by randomly resampling the original data with replacement. For each bootstrap sample, we computed the statistic of interest:
\begin{itemize}
    \item The sample mean $\bar{x}$
    \item The sample standard deviation $s$, which is the square root of Eq.~\ref{eq:s}.
\end{itemize}

From the bootstrap distribution of each statistic, we constructed 95\% confidence intervals using the \textit{bias-corrected and accelerated (BCa)} method \cite{diciccio1996bootstrap}. Unlike the simpler percentile method, which takes the 2.5th and 97.5th percentiles directly from the bootstrap distribution, the BCa method adjusts these cutoffs to correct for potential bias and skewness in the bootstrap estimates. This adjustment improves the accuracy and reliability of the resulting confidence intervals, especially for nonlinear statistics such as the standard deviation.

\subsection{Normality Assessment}

Before conducting inferential statistical analyses, we needed to test whether the sample data we used could reasonably be assumed to come from a normally distributed population. This assumption is critical for the validity of parametric procedures, such as CI estimation using the $t$-distribution or chi-square distribution. Thus, we applied the Shapiro-Wilk test to each of our samples to assess normality before proceeding with the calculation of the CIs. 

The Shapiro-Wilk test is a powerful statistical test to assess the normality of a dataset, particularly effective for small samples (generally $n$ less than 50)\cite{statistics} such as the samples used in this study, each consisting of 10 observations. The test evaluates whether a sample came from a normally distributed population by comparing ordered sample values against the expected values of a normal distribution. The test statistic $W$ is calculated as:
\begin{equation}
W = \frac{\left(\sum_{i=1}^{n} a_i x_{(i)}\right)^2}{\sum_{i=1}^{n} (x_i - \bar{x})^2},
\end{equation}
where $x_{(i)}$ represents the ordered sample values (from smallest to largest), $a_i$ are weights derived from the mean, variance and covariance of the order statistics from a standard normal distribution and $\bar{x}$ is the sample mean.

The weights $a_i$ are computed as:
\begin{equation}
\mathbf{a} = \frac{\mathbf{m}^T\mathbf{V}^{-1}}{(\mathbf{m}^T\mathbf{V}^{-1}\mathbf{V}^{-1}\mathbf{m})^{1/2}},
\end{equation}
where the vector $\mathbf{m}$ represents the expected values of standard normal order statistics, and $\mathbf{V}$ is their covariance matrix. 

\begin{figure}[t]
  \centering
  \includegraphics[width=0.92\linewidth]{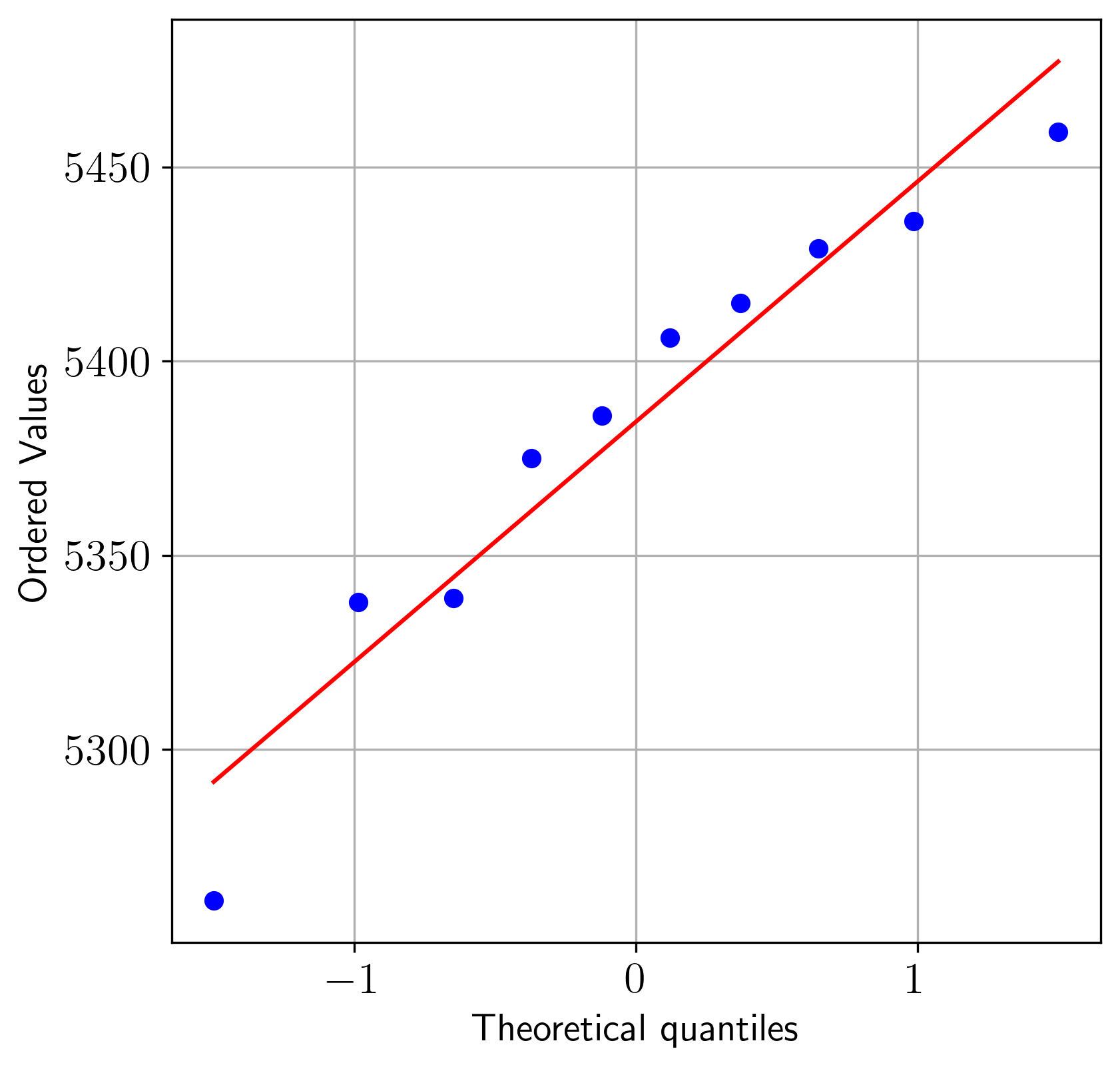}
  \caption{Q-Q plot of the sample data for the number of injections detected at FAR = 100/month, for the datasets analyzed in Sec.\ref{sec:same_noise_diff_inj}.} 
  \label{fig:qq}
\end{figure}

The $W$ statistic is bounded between 0 and 1, with values closer to 1 indicating that the data follow a normal distribution. For small samples, the test's critical values have been tabulated. The null hypothesis (that the sample comes from a normally distributed population) is rejected if $W$ is below the critical value at a given significance level, typically 0.05. The Shapiro-Wilk test maintains excellent power to detect deviations from normality even with limited data points, making it the preferred normality test for small samples in statistical analysis.

For a more comprehensive analysis as well as a visual confirmation of the normality of our data, we also generated Q-Q (quantile-quantile) plots like the one in Fig. \ref{fig:qq} for all our samples. A Q-Q plot compares the quantiles of the sample data to the quantiles of a standard normal distribution. If the data follows a normal distribution, the points should approximately align along a straight line. Deviations from this line, particularly in the tails, indicate deviations from normality. The Q-Q plots for most of our samples showed that the data closely followed the theoretical normal distribution, further supporting the assumption of normality, as confirmed by the Shapiro-Wilk test results. However, only the Q-Q plot for the case where normality was rejected is presented.

\section{Sensitive Distance}
\label{appendix:sensitive_distance}

In GW astronomy, a core metric for evaluating and comparing the sensitivity of detection pipelines is the \textit{sensitive volume}. This metric represents the effective three-dimensional region of space within which signals can be confidently detected at a given false alarm rate $\mathcal{F}$. Rather than quoting this volume directly, it is common to report the \textit{sensitive distance}, defined as the radius of a sphere with the equivalent volume.

To calculate the sensitive distance, we need first to evaluate the sensitive volume. In practice, we find the sensitive volume by adding a large number $N$ of simulated signals to the data and measuring the pipeline's ability to identify them. If the simulated signals are drawn uniformly in volume up to a maximum distance \(d_{\mathrm{max}}\), then the sensitive volume can be approximated as described in \cite{usman2016pycbc, challenge1}:
\begin{equation}
    V(\mathcal{F}) = V_{\mathrm{max}} \frac{N_{\rm inj, \mathcal{F}}}{N} \label{C.1}
\end{equation}
\noindent where $V_{\mathrm{max}}$ is the volume of a sphere with radius $d_{\mathrm{max}}$, and $N_{\rm inj,\mathcal{F}}$ is the number of found injections at a FAR of $\mathcal{F}$. An injection is considered ``found'' if there exists at least one foreground event that is within $\pm \Delta t$ of the injection coalescence time $t_c$, where $\Delta t$ is the timing uncertainty assigned by the search algorithm. The count $N_{\rm inj, \mathcal{F}}$ includes only those foreground events whose ranking statistic exceeds the threshold corresponding to the given FAR \cite{challenge1}.

Distributing signals uniformly in volume leads to a majority of signals being injected at large distances. In the case of low-mass signals, this results in low SNR values. As a consequence, only a small fraction of the total injections can be detectable. Since the total number of injections is limited, the above choice leads to poor sampling in the region where signals transition from undetectable to detectable. To mitigate this, instead of sampling uniformly in volume, we sample over the chirp distance. The chirp distance is defined as
\begin{equation}
    d_c = d\left(\frac{\mathcal{M}_{c,0}}{\mathcal{M}}\right)^{5/6} \label{C.2}
\end{equation}
where $d$ is the luminosity distance and $\mathcal{M}_{c,0}$ is the fiducial chirp mass used as a reference. From equation~\ref{C.2}
one can see that by injecting uniformly in $d_c$, the maximum luminosity distance at which an injection is placed depends on its $\mathcal{M}$. This mass-dependent scaling effectively increases the number of detectable low-mass systems. 

Since $d_{\mathrm{max}}$ varies with $\mathcal{M}$, each source with chirp mass $ \mathcal{M}_{c,i}$ is instead injected uniformly within a sphere of volume $V_i \propto d_{\mathrm{max},i}^3 \propto \mathcal{M}_i^{5/2}$. As a result, each detected injection must be reweighted by the effective volume it represents. Therefore, equation~\ref{C.1} becomes: 
\begin{equation}
    V(\mathcal{F}) = \frac{V(d_{\mathrm{max}})}{N} \sum\limits_{i=1}^{N_{\mathrm{inj},\mathcal{F}}}\left(\frac{M_{c,i}}{M_{\mathrm{max}}}\right)^{5/2} \label{C.3}
\end{equation}
\noindent where $\mathcal{M}_{c,i}$ is the chirp mass of the $i$-th found injection with FAR $\mathcal{F}$, and $d_{c, \mathrm{max}}$, $\mathcal{M}_{c, \mathrm{max}}$ is the maximum chirp distance and chirp mass from the collection of signals that were injected in the data \cite{challenge1}. The sensitive distance can then be calculated from
\begin{equation}
    S(\mathcal{F}) = \left(\frac{3}{4\pi} V(\mathcal{F})\right)^{1/3}. \label{C.4}
\end{equation}

\bibliography{AresGW_comparison}

\end{document}